\documentclass{aa}  
\usepackage{graphicx}
\usepackage[colorlinks=true,citecolor=blue]{hyperref}
\usepackage{multirow}

\usepackage{wasysym}

\usepackage{xcolor}

\usepackage{comment}
\usepackage{float}
\usepackage{dblfloatfix}
\usepackage{multicol,multirow}

\usepackage{amsmath}

\usepackage{microtype}
\microtypesetup{protrusion=true,expansion=true}

\usepackage{txfonts}
\defcitealias{Valles-Perez_2025_accr-i}{Paper I}

\begin{document}

   \title{The eventful life journey of galaxy clusters}
   \subtitle{II. Impact of mass accretion on the thermodynamical structure of the ICM}

   \titlerunning{Mass accretion and the thermodynamical profiles of the ICM}

   \author{David Vallés-Pérez
          \inst{1,2}\fnmsep\thanks{\email{david.vallesperez@unibo.it}.}
          \and
          Susana Planelles\inst{2,3}
          \and 
          Vicent Quilis\inst{2,3}
          }

   \institute{Dipartimento di Fisica e Astronomia, Università di Bologna, Via Piero Gobetti 93/2, IT-40129 Bologna, Italy
              \and
              Departament d’Astronomia i Astrofísica, Universitat de València, ES-46100 Burjassot (València), Spain
              \and 
              Observatori Astronòmic, Universitat de València, ES-46980 Paterna (València), Spain}

   \date{\today}

  \abstract
   {The internal structure of the intracluster medium (ICM) is tightly linked to the assembly history and physical processes in groups and clusters, but the role of recent accretion in shaping these profiles has not been fully explored.}
   {We investigate to what extent mass accretion accounts for the variability in ICM density and thermodynamic profiles, and what can present-day structures reveal about their formation histories.}
   {We analyse a hydrodynamical cosmological simulation including gas cooling but no feedback, to isolate the effects of heating from structure formation. Median profiles of ICM quantities are introduced as a robust description of the bulk ICM. We then examine correlations between mass accretion rates or assembly indicators with the profiles of temperature, entropy, pressure, gas and dark-matter density, as well as their scatter.}
   {Accretion in the last dynamical time strongly lowers central gas densities, while leaving dark matter largely unaffected, producing a distinct signature in the baryon depletion function. Pressure and entropy show the clearest dependence on accretion, whereas temperature is less sensitive. The radii of steepest entropy, temperature, and pressure shift inward by $\sim(10-20)\%$ between high- and low-accretion subsamples. Assembly-state indicators are also related to the location of these features, and accretion correlates with  the parameters of common fitting functions for density, pressure, and entropy.}
   {Recent accretion leaves measurable imprints on the ICM structure, highlighting the potential of thermodynamic profiles as diagnostics of cluster growth history.}


   \maketitle

%-------------------------------------------------------------------
\section{Introduction}
\label{s:intro}
%--------------------------------------------------------------------

Most of the baryons in galaxy clusters reside in a hot, diffuse phase known as the intracluster medium (ICM), with typical number densities ranging ${n \sim 10^{[-5, \, -1]}\, \mathrm{cm}^{-3}}$. Together with the even more diffuse warm-hot intergalactic medium (WHIM), they extend out to several virial radii \citep{Dave_2001, Walker_2019}. As their dominant baryonic component, the ICM\footnote{Hereafter, we use the term ICM to refer to all diffuse baryons bound to a galaxy cluster, i.e., also containing the WHIM.} is central to our understanding of galaxy clusters, the largest virialised systems in the Universe, as it provides a record of the gravitational and non-gravitational processes at play during cluster assembly (e.g. \citealp{Kravtsov_2012, Planelles_2015}).

As the diffuse gas from the intergalactic medium falls into the dark matter (DM) halo potential well, the ICM emerges as an almost fully-ionised plasma and reaches temperatures of ${\sim 10^{[7,8]} \, \mathrm{K}}$ due to adiabatic and shock compression \citep{Evrard_1990, Quilis_1998}. The dissipation of turbulent motions, naturally arising during hierarchical assembly \citep{Dolag_2005, Vazza_2017}, together with a plethora of non-gravitational processes (e.g. feedback from star formation and active galactic nuclei, \citealp{McNamara_2007}, as well as microphysical transport processes, \citealp{Walker_2019}) all alter the properties and distribution of diffuse baryons. Altogether, these mechanisms determine the radial structure of fundamental ICM quantities, such as density, temperature, entropy, and pressure.

The thermal ICM can be observed directly in the X-ray band, where its continuum (bremsstrahlung) and line emission inform about gas density, temperature and metallicity \citep{Bohringer_2010}. Complementary information can be obtained in the microwave band from the \citet[][SZ]{Sunyaev_1972} effect, which probes the integrated pressure of the hot electrons \citep{Mroczkowski_2019}. Large campaigns in both these bands (either separately, \citealp{Merloni_2024, ACTDR6}; or in combination, \citealp{Eckert_2019}) have provided constraints on the average profiles, their intrinsic scatter, etc. Near-to-mid future instruments (\textsc{newAthena}, \citealp{Cruise_2025}; \textsc{Simons Observatory}, \citealp{Abitbol_2025}) will likely expand the radial range towards cluster outskirts, and increase the sample sizes, reaching higher redshifts.

In this context, an accurate theoretical modelling is crucial to take the most from current and future observational efforts. At its most basic, clusters have been analytically described within the self-similar model \citep{Kaiser_1986} with simple rescaling prescriptions. However, departures from self-similarity have been both predicted from simulations \citep{Borgani_2004} and observed \citep{Arnaud_2005} for decades. Clusters are dynamically young objects, with very diverse assembly histories (\citealp{Wong_2012, Valles-Perez_2020}; \citealp{Valles-Perez_2025_accr-i}, hereon \citetalias{Valles-Perez_2025_accr-i}), environments \citep{Kuchner_2020} and internal processes \citep[e.g.][]{LeBrun_2014}, what imprints a substantial variability in their internal structure and thermodynamical state.

Within this picture, simulations --which allow us to track the whole evolution of these systems-- become essential tools to link the assembly history of clusters to their internal structure. 
In particular, several studies have sought to constrain the baryon distribution under different feedback models and across a range of radial apertures \citep{Planelles_2013, Battaglia_2013, Angelinelli_2022, Rasia_2025}, while also looking at how accretion and mergers shape the gas fraction (and, in turn, X-ray detectability, \citealp{Marini_2025}). Regarding ICM thermodynamics, \citet{Hernandez-Martinez_2025} recently studied the thermodynamical profiles of local cluster analogues in relation to their assembly histories and cool-coredness (also closely related to dynamical state). Bridging the gap with observations, \citet{Biffi_2013} investigated the relation between the kinematic state of the ICM (as a proxy for dynamical state) and X-ray luminosity, while \citet{DeLuca_2021} examined the complex relation of X-ray and SZ morphology (including, for instance, the radial scatter and concentration), and three-dimensional measures of cluster assembly.

Perhaps the study most directly addressing the connection between ICM thermodynamics and cluster assembly is \citet{Lau_2015}, who looked at the impact of mass accretion (quantified from the radial velocity near the virial radius, $R_\mathrm{vir}$) in a non-radiative simulation on the pressure, temperature and entropy profiles, with a special focus on cluster outskirts. The main trend reported by the authors amounts to a shift in the location of the accretion shock, with stronger accretion pushing this feature inwards. In contrast, the impact on central regions is reduced. This likely stems from the absence of cooling mechanisms, which render heating due to structure formation processes less important. A more recent analysis, based on the \textsc{Flamingo} simulations, has also explored how the density and temperature profiles of the hot gas vary with halo accretion rates, although with a broader focus on feedback effects \citep{CorreaMagnus_2025}.

Building upon the previous literature, in this work we aim to establish what is the effect of cluster mass growth, either from mergers or from smooth accretion, on the ICM profiles across their whole radial extent. To do so, we analyse a cosmological hydrodynamical simulation accounting for gas cooling, but no feedback mechanisms, in such a way that we can isolate the heating due to structure formation processes. In particular, we aim to assess: \textit{(i)} the impact of selecting clusters by their recent accretion rate on their internal structure, \textit{(ii)} the extent to which accretion explains the deviations from self-similarity of the profiles, \textit{(iii)} whether commonly-used assembly state indicators constrain the thermodynamical profiles, and \textit{(iv)} besides the central tendency of the profiles, what can be said about their intrinsic scatter.

The manuscript is organised as follows. In Sect. \ref{s:methods} we introduce the simulation setup and the main analysis techniques. The results are presented in several subsections within Sect. \ref{s:results}. Finally, we further discuss and put our results into context within the literature in Sect. \ref{s:conclusions}. Appendices \ref{app:self-similar}-\ref{app:fits} contain several additional tests and details.

%--------------------------------------------------------------------
\section{Methods}
\label{s:methods}
%--------------------------------------------------------------------

Here, we present the key methodological aspects of this study, beginning in Sect. \ref{s:methods.simulation} with our simulation and cluster catalogue. The method for extracting radial profiles and stacking them are then covered, respectively, in Sects. \ref{s:methods.profile} and \ref{s:methods.stacking}.

%----------------------
\subsection{The simulation and cluster catalogue}
\label{s:methods.simulation}
%----------------------
The present analyses are based on a $\Lambda$CDM $N$-Body+hydrodynamical simulation, previously employed in earlier studies (e.g., \citealp{Valles-Perez_2024_NatAstro} and \citetalias{Valles-Perez_2025_accr-i}). For completeness, we provide here a succinct description of the simulation details, referring the reader to those references for further discussion. 

The simulation tracks the evolution of a periodic cube of comoving side $L=100 \, h^{-1} \, \mathrm{Mpc}$, in a flat $\Lambda$CDM cosmology with matter density ${\Omega_m=0.31}$, baryon fraction $f_b = 0.155$, Hubble dimensionless parameter $h=0.678$, and primordial power spectrum given by the index $n_s=0.96$ and the normalisation $\sigma_8=0.82$, consistent with \citet{Planck_2020} constraints. The initial conditions, set at $z_\mathrm{ini}=100$, are evolved in time with the adaptive mesh refinement (AMR) code \textsc{masclet} \citep{Quilis_2004} accounting for gravitational and hydrodynamical forces, as well as gas cooling. No star formation nor feedback are accounted for in this run. The deliberate choice of this set-up was motivated by the fact that, while it may produce unrealistic thermodynamical distributions in cluster cores due to excessive, unopposed cooling, it serves us to isolate the effect of cluster assembly (gas accretion and mergers) on the thermodynamical state of the ICM, without confusion due to other (non-gravitational) heating feedback mechanisms. The coarsest and peak spatial resolutions are $\Delta x_\mathrm{base} = 576 \, \mathrm{kpc}$ and $\Delta x_\mathrm{peak} = 9 \, \mathrm{kpc}$, where refinement is triggered by overdensities and converging flows. Base and peak DM mass resolutions are $M_\mathrm{DM,base} = 7.7 \times 10^9 \, M_\odot$ and $M_\mathrm{DM,peak} = 1.5 \times 10^7 \, M_\odot$. The effective refinement is such that $\gtrsim 50\%$ of the mass within $R_\mathrm{vir}$ of our cluster sample is resolved at $\Delta x_5 = 18 \, \mathrm{kpc}$ resolution at $z=0$ or better.\footnote{In terms of volume, $\gtrsim 75\%$ of the virial volume is resolved at $\Delta x_4 = 36 \, \mathrm{kpc}$ or better.}

Cluster catalogues are extracted from the DM distribution with the public halo finder \textsc{ASOHF} \citep{Planelles_2010, Knebe_2011, Valles-Perez_2022}. For all purposes within this work, the density peak of the DM distribution is chosen as the centre of the cluster. The basic sample, constructed at $z=0$, consists of 31 clusters (above $M_\mathrm{DM}^{z=0} > 10^{14} \, M_\odot$) and 358 groups ($M_\mathrm{DM}^{z=0} > 10^{13} \, M_\odot$). As we assume self-similarity to rescale the profiles (see Sect. \ref{s:methods.stacking} below), unless stated otherwise, throughout this article we will refer to the whole sample as \textit{clusters} to ease notation, even though it comprises clusters and groups.

Table \ref{tab:mass_sample} summarizes in greater detail the mass composition of the selected systems, as well as their hot gas. The median gas fraction at $R_{200m}$, over the whole sample, lies around $(80-85)\%$ of the cosmic one, which is consistent with the baryon fractions for $\sim 10^{14} \, M_\odot$ clusters as reported by, e.g., \citet{Angelinelli_2022} using the \textsc{Magneticum} suite. At lower masses (which are the ones dominating our sample), this baryon fraction in sensibly higher than the values reported in the literature, which range around $(60-70)\%$ of the cosmic value, as a consequence of overcooling. 

While the present analysis is restricted at $z=0$, merger trees obtained from tracking the most bound particles are used to compute the accretion rates $\Gamma_{200m} = \mathrm{d} \log M_{200m} / \mathrm{d} \log a$ over the last dynamical time ($\tau_\mathrm{dyn}$) as discussed in \citetalias{Valles-Perez_2025_accr-i}.

\begin{table*}[]
    \centering
    \caption{Summary statistics accounting for the mass distribution of our group and cluster sample.}
    \begin{tabular}{c|cccccccc}
       & $\langle X \rangle$ & $X_\mathrm{min}$ & $X_{2.5\%}$ & $X_{16\%}$ & $X_{50\%}$ & $X_{84\%}$ & $X_{97.5\%}$ & $X_\mathrm{max}$  \\ \hline
       $\log_{10} ( M_{200m}^\mathrm{DM} / M_\odot )$ & $13.48$ & $13.02$ & $13.07$ & $13.15$ & $13.42$ & $13.81$ & $14.30$ & $14.74$ \\
       $\log_{10} ( M_{200m}^\mathrm{gas} / M_\odot )$  & $12.65$ & $12.06$ & $12.20$ & $12.31$ & $12.58$ & $12.99$ & $13.49$ & $13.95$ \\
       $\log_{10} ( M_{200m}^\mathrm{tot} / M_\odot )$  & $13.54$ & $13.08$ & $13.13$ & $13.20$ & $13.48$ & $13.87$ & $14.36$ & $14.81$ \\
       $f_\mathrm{gas}^{200m}$  & $0.128$ & $0.091$ & $0.107$ & $0.119$ & $0.129$ & $0.137$ & $0.147$ & $0.162$ \\ 
       \hline 
       $\log_{10} ( M_{200m}^\mathrm{hot \;gas} / M_\odot )$ & $12.59$ & $11.97$ & $12.06$ & $12.23$ & $12.53$ & $12.96$ & $13.45$ & $13.93$ \\
       $f_\mathrm{hot \; gas}^{200m}$ & $0.112$ & $0.066$ & $0.078$ & $0.093$ & $0.113$ & $0.130$ & $0.141$ & $0.162$ \\
       $f_\mathrm{hot \; gas}^{200m} / f_\mathrm{gas}^{200m}$ & $0.875$ & $0.521$ & $0.620$ & $0.742$ & $0.899$ & $0.995$ & $1.000$ & $1.000$ 
    \end{tabular}
    \tablefoot{
    Each row in the first block corresponds to a different quantity, namely the DM, gas and total masses within $R_{200m}$ (in logarithmic scale), and the baryon fraction within the same aperture. The second block contains the information relative to the hot gas (which we define with a temperature threshold of $T_\mathrm{thr} = 10^6 \, \mathrm{K}$, similar to, e.g., \citealp{Rasia_2025}). Columns present the arithmetic mean $\langle X \rangle$, minimum $X_\mathrm{min}$, several percentiles $X_{p\%}$, and maximum $X_\mathrm{max}$ of each quantity $X$.
    }
    \label{tab:mass_sample}
\end{table*}

%----------------------
\subsection{Profile making}
\label{s:methods.profile}
%----------------------

Most customarily, radial profiles of additive (i.e., extensive; e.g., mass) quantities are obtained by summing over the resolution elements (either cells or particles) in pre-defined radial shells and normalising by the integration volume to define the corresponding density. For intensive quantities (e.g., temperature), the values are instead computed by averaging over the elements in the shell, possibly allowing for a weighting function (e.g., mass, emission, etc.).

This direct procedure has, however, a number of drawbacks. First, the finite width of the shells needs to be carefully tuned, since shells that are too wide would smear out the profile, and could be dominated by extreme values, especially when slopes are steep. Conversely, shells that are too narrow are noisy due to low-number statistics. Secondly, when considering mean profiles, the effect of clumps or very anisotropic environments (the latter, especially in cluster outskirts) dominates the value of the profiles. However, these contributions do not reflect the bulk ICM, since the dynamics of these structures (accounting for a significant fraction of cluster baryons, but a negligible volume; see, e.g., \citealp{Angelinelli_2021}) are usually decoupled from the smooth, volume-filling gas. While the first caveat is usually corrected by carefully tuning the binning (which in turn depends on resolution), the second needs ad-hoc clump and substructure-removal steps, leading to resulting profiles that depend continuously on the free parameters of these cleaning procedures (e.g., \citealp{Zhuravleva_2013}).

In observations, it has become fairly extended to employ the azimuthal median technique (\citealp{Eckert_2015}; and subsequent works, e.g., \citealp{Ghiradini_2019}; also studied in the context of projected profiles from simulations, e.g. \citealp{Ansarifard_2020, Towler_2023}) to alleviate the emissivity bias due to strongly overdense clumps and recover the thermodynamical state of the smooth, volume-filling ICM\footnote{Together with other corrections, often informed by simulations \citep[e.g.][]{Vazza_2013, Planelles_2017}, to correct for the unresolved clumping.}. This was partly motivated by \citet{Zhuravleva_2013}, who quantified the properties of clumps in simulated ICMs and already argue that median profiles can be a better representation of the bulk ICM properties. Nevertheless, in three-dimensional profiles from simulations, the predominant practice is still to use mean profiles, most often with some substructure-excision procedure.

In contrast to the more common radial binning methods, in this work we adopt an alternative procedure to extract three-dimensional profiles by computing, at each radius $r$, the median value of the quantities of interest ($f$) over $N_\theta \times N_\phi$ angular directions, equally spaced in the unit sphere,
\begin{equation}
    f(r) = \mathop{\mathrm{med}}_{\cos \theta, \phi} \; f(r, \theta, \phi).
\end{equation}

To determine $f(r, \theta, \phi)$, we evaluate the quantity of interest at the Cartesian coordinates corresponding to $(r, \theta, \phi)$, by trilinear interpolation among the eight neighbouring cells at the highest available level of the AMR hierarchy, limited so that the local cell size does not fall below the radial spacing $\Delta r$. These profiles naturally account for the volume-filling, hot ICM, without the need of imposing any temperature threshold or explicit substructure removal step. Compared to the median profiles of \citet{Zhuravleva_2013}, this procedure has the additional advantage of being insensitive to any radial bin width. We illustrate the properties of these profiles, and compare them to mean and mode profiles, in Sect. \ref{s:results.mean_vs_median}.

%----------------------
\subsection{Profile stacking}
\label{s:methods.stacking}
%----------------------
Stacked profiles (used in Sects. \ref{s:results.density}, \ref{s:results.thermodynamic} and \ref{s:results.fits}) of a given quantity, $X(r)$, over a given sample are computed by: 
\begin{enumerate}
    \item Obtaining the individual profiles of the quantity of interest, $X(r)$, from\footnote{$R_{\Delta_m}$ is the radius encompassing an overdensity $\Delta_m$ with respect to the mean matter density.} $r_\mathrm{min}=0.01 R_{200m}$ to $r_\mathrm{max} = 4 R_{200m}$, with $\Delta \log r = 0.01 \, \mathrm{dex}$.
    \item Rescaling the individual profiles to ${\mathcal{X}(\mathcal{R}) \equiv {X(\mathcal{R} \cdot R_\mathrm{norm}) / X_\mathrm{norm}}}$, where we take $\mathcal{R}$ to be the radius in units of ${R_\mathrm{norm}=R_{200m}}$ and the normalisation constants $X_\mathrm{norm}$ correspond to the self-similar values specified in App. \ref{app:self-similar}.
    \item Resampling the profiles to be computed at the same values of $\mathcal{R}$. This is achieved by linear interpolation in double-logarithmic space.
    \item Stacking the profiles by computing, at each $\mathcal{R}$, the biweight robust mean \citep[e.g.][]{Beers_1990} of the values of $\log \mathcal{X}$.
\end{enumerate}

Typical uncertainties around the stacked profiles are determined as the $16$-th and $84$-th percentiles at each $\mathcal{R}$ over ${N_\mathrm{boots} = 100}$ bootstrap resamplings. 

Finally, the logarithmic derivatives of the profiles are computed using \citet{Savitzky_1964} cubic filters over a window length of 21 points (corresponding to $0.21 \, \mathrm{dex}$ in radius).

%--------------------------------------------------------------------
\section{Results}
\label{s:results}
%--------------------------------------------------------------------

We start the presentation of our results with a comparison among different solid-angle averaging methods for building radial profiles (Sect. \ref{s:results.mean_vs_median}), and argue why the median profiles are the optimal choice for representing the smooth ICM. Using these, in Sects. \ref{s:results.density} and \ref{s:results.thermodynamic} we discuss the impact of cluster assembly on structural (i.e. density) and thermodynamical profiles of the ICM. In Sect. \ref{s:results.indicators} we explore the effect of selecting objects by single, cluster-wide properties on the profiles, while in Sect. \ref{s:results.scatter} we discuss the angular scatter around individual profiles and the self-similarity among different clusters. Finally, we consider the impact of accretion on the estimation of parameters from widely used fitting forms in Sect. \ref{s:results.fits}.

%----------------------
\subsection{Mean and median profiles}
\label{s:results.mean_vs_median}
%----------------------

\begin{figure*}
    \centering
    \includegraphics[width=0.5\textwidth]{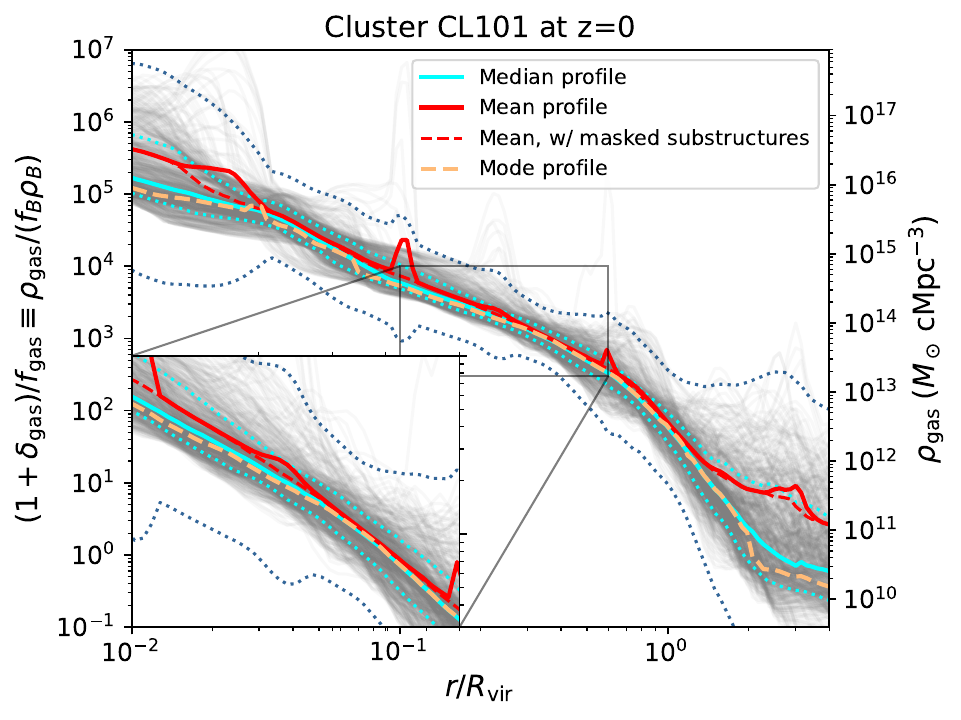}~
    \includegraphics[width=0.5\textwidth]{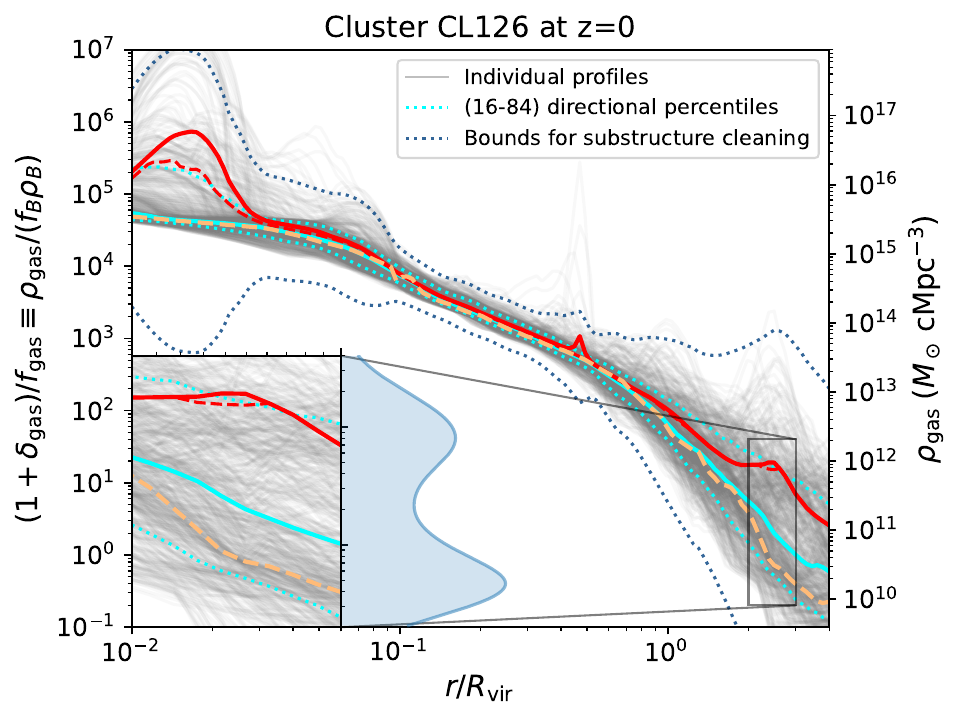}
    \caption{Comparison between mean (red solid line), median (cyan solid line) and mode (orange dashed line) gas density profiles in two galaxy clusters (corresponding to the two panels). In addition to the standard mean profiles, also profiles with substructures having been masked are shown as the dashed, red lines. Light gray lines indicate the individual directional profiles, while cyan dotted lines are the $(16-84)\%$ percentiles at each $r$ to give a better idea of the width of the density distribution. Dark blue dotted lines indicate the thresholds for the substructure cleaning algorithm from which the red dashed line has been obtained. The insets provide a zoomed-in view of two regions to highlight the differences among the profiles. The blue filled line at the right of the right-hand panel inset depicts the distribution of values of density at $r \sim 3 R_\mathrm{vir}$, to highlight its bimodality.}
    \label{fig:median_vs_mean_density}
\end{figure*}

In Sect. \ref{s:methods.profile}, we have introduced a novel approach for computing radial profiles, aimed at ameliorating the impact of clumps, substructures and anisotropic environments, under the assumption of it being a better representation of the bulk ICM. In this section we briefly discuss its differences from standard mean profiles, by highlighting the advantages of our median profiles of gas density in two clusters at $z=0$ (Fig. \ref{fig:median_vs_mean_density}).

Focusing on CL101 (left-hand side panel), we show with gray lines the individual directional profiles, while cyan (red) solid lines indicate the median (mean) radial profiles. For this comparison, mean profiles are computed as the arithmetic mean of $N_\theta \times N_\phi = 25 \times 25$ directional profiles (instead of summing in a radial bin), so the problematics of finite-width shells are suppressed. There are three clearly distinct regions. 

\begin{itemize}
    \item In the innermost regions ($r/R_\mathrm{vir} \lesssim 0.04$)\footnote{Here, $R_\mathrm{vir}$ is computed according to the \citet{Bryan_1998} prescription.}, mean profiles exceed median profiles by over a factor of $2$ due to the very high-density tails reaching $\Delta_\mathrm{gas} = \rho_\mathrm{gas} / \rho_B > 10^7$. This is not mitigated by the substructure-excised profiles (dashed red line, removing cells above $3.5 \sigma$ in $\log \rho$ at each $r$; \citealp{Zhuravleva_2013}) due to the broad density distribution (see dashed, dark blue lines, indicating the thresholds for clump excision).
    \item At intermediate radii ($0.04 \lesssim r/R_\mathrm{vir} \lesssim 1.5$), where profiles tend to be more self-similar, the differences between the mean and median profiles are greatly reduced, with the exception of some bumps in the mean profile associated with prominent clumps that can be filtered out by the excision procedure. However, the inset presents a zoomed-in view showing how mean profiles overshoot the median one by $(10-20)\%$ even in this region of regularity.
    \item In the outskirts ($r/R_\mathrm{vir} \gtrsim 1.5$), mean and median profiles diverge again by almost an order of magnitude. In this region, besides the impact of clumping, the very anisotropic environment drives mean densities up even if only a small portion of the solid angle points in the direction of other overdense structures (filaments, other groups and clusters, etc.). In this sense, the median profiles are less sensitive to the environment anisotropy and suppress this two-halo contribution \citep{Mo_1996, Sheth_2001} more effectively.
\end{itemize}

The figures also show the mode profiles, i.e., the most likely value of gas density for each $r$, obtained from the directional profiles using a kernel density estimate (dashed, orange line). This profile could arguably be the best representation of the smooth ICM and is robust under clumping. Our median profiles follow closely the mode profiles in the inner and intermediate region, where the former only exceed the latter by $\sim 5\%$.

The other cluster (CL126; right-hand side panel) follows similar patterns, but exemplifies an important drawback of the mode profile, namely its behaviour in radial regions where the density probability distribution may not be monomodal. This is especially frequent in cluster outskirts. The inset shows a closer look at the $2 < r/R_\mathrm{vir} < 3$ radial range, where the density of gray lines (individual directional profiles) already hints two regions of higher concentration. This is shown more clearly at the right of the inset, where we plot a histogram with the distribution of values of density, highlighting its bimodality. In such a scenario, mean profiles are dominated by the high-density region, while mode profiles are very unstable and may switch from one branch to the other depending on slight changes on the solid angle corresponding to a denser (e.g. filaments or a bridge with a nearby cluster) or a lower-density (i.e. voids) environments.\footnote{This bimodality may be also contributed by the asphericity of the accretion shock (e.g., \citealp{Valles-Perez_2024_NatAstro}, their figure S1). Since the accretion shock produces a density jump of a factor of $\sim 4$, but it may be located at different radii in each direction, at any $r$ the value of the mode profile will be extremely sensitive to the distribution of distances to the shock shell.} %Similar features, associated to very anisotropic density distributions, are also found, to a smaller extent, in the inner regions of CL101 (left-hand side panel).

The considerations above lead us to choose the median profile as a conservative representation of the bulk, smooth component of the ICM for the remainder of this work.
%----------------------
\subsection{Density profiles}
\label{s:results.density}
%----------------------

\begin{figure*}
    \centering
    \includegraphics[width=0.33\textwidth]{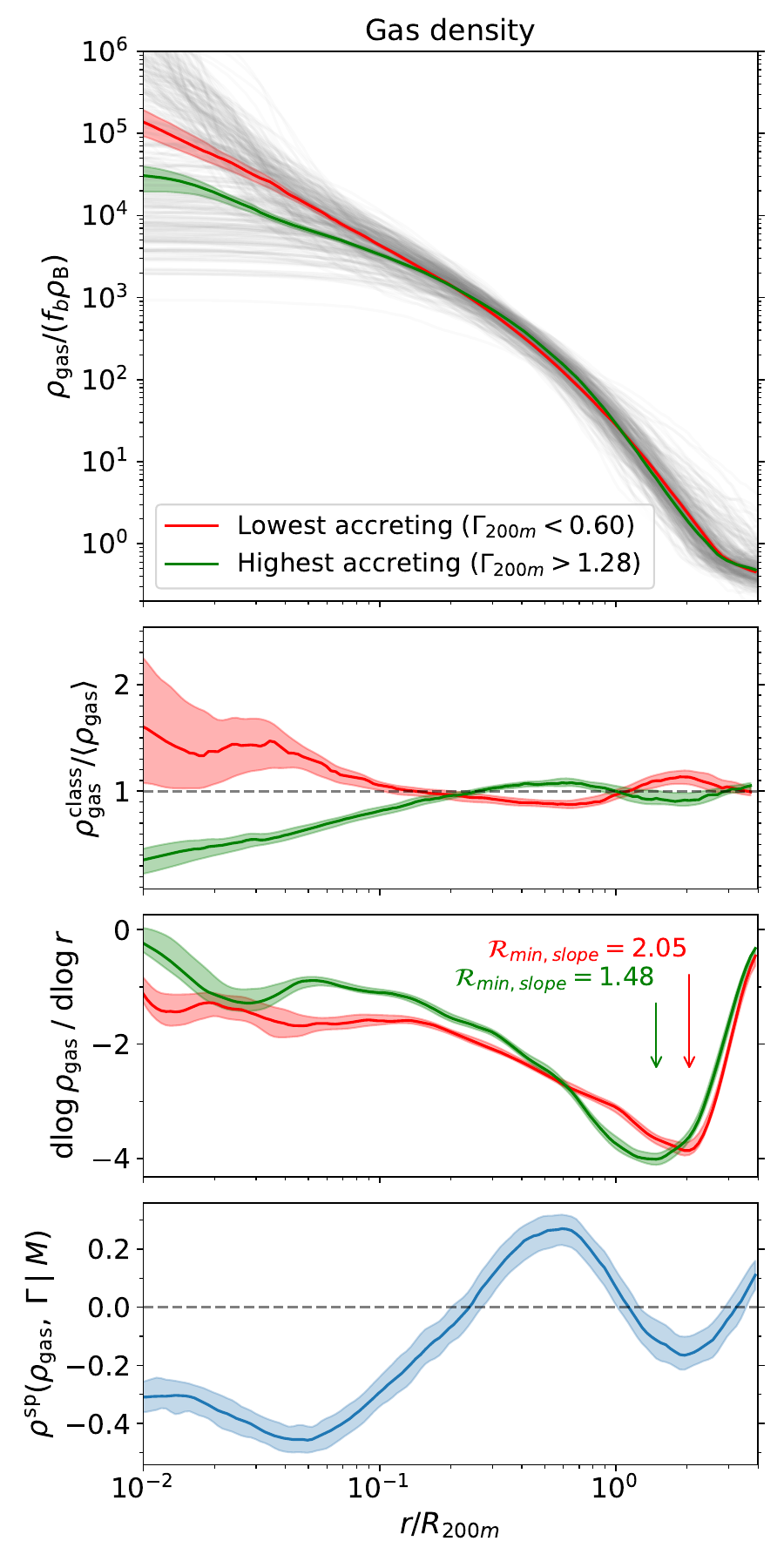}~
    \includegraphics[width=0.33\textwidth]{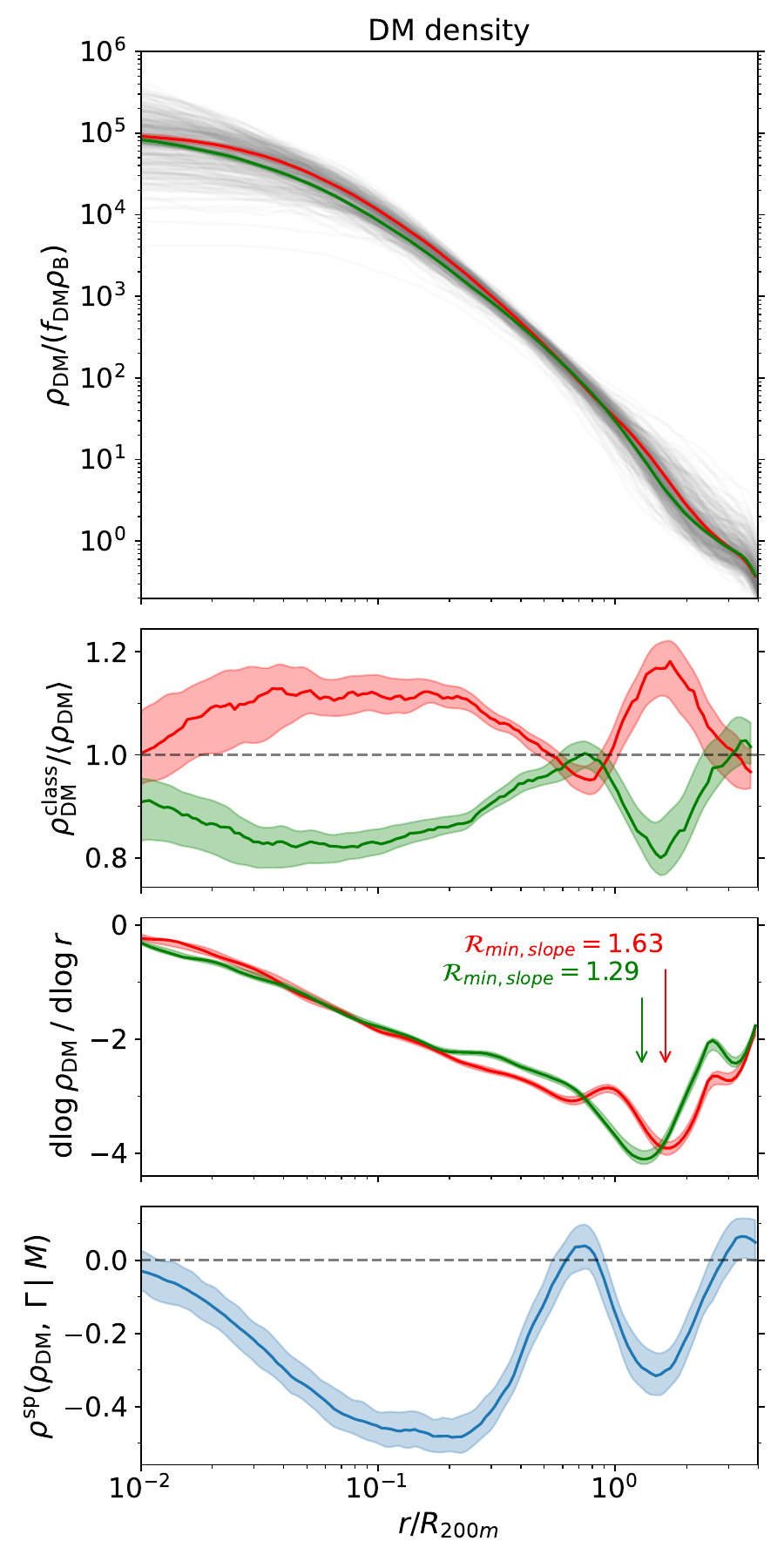}~
    \includegraphics[width=0.33\textwidth]{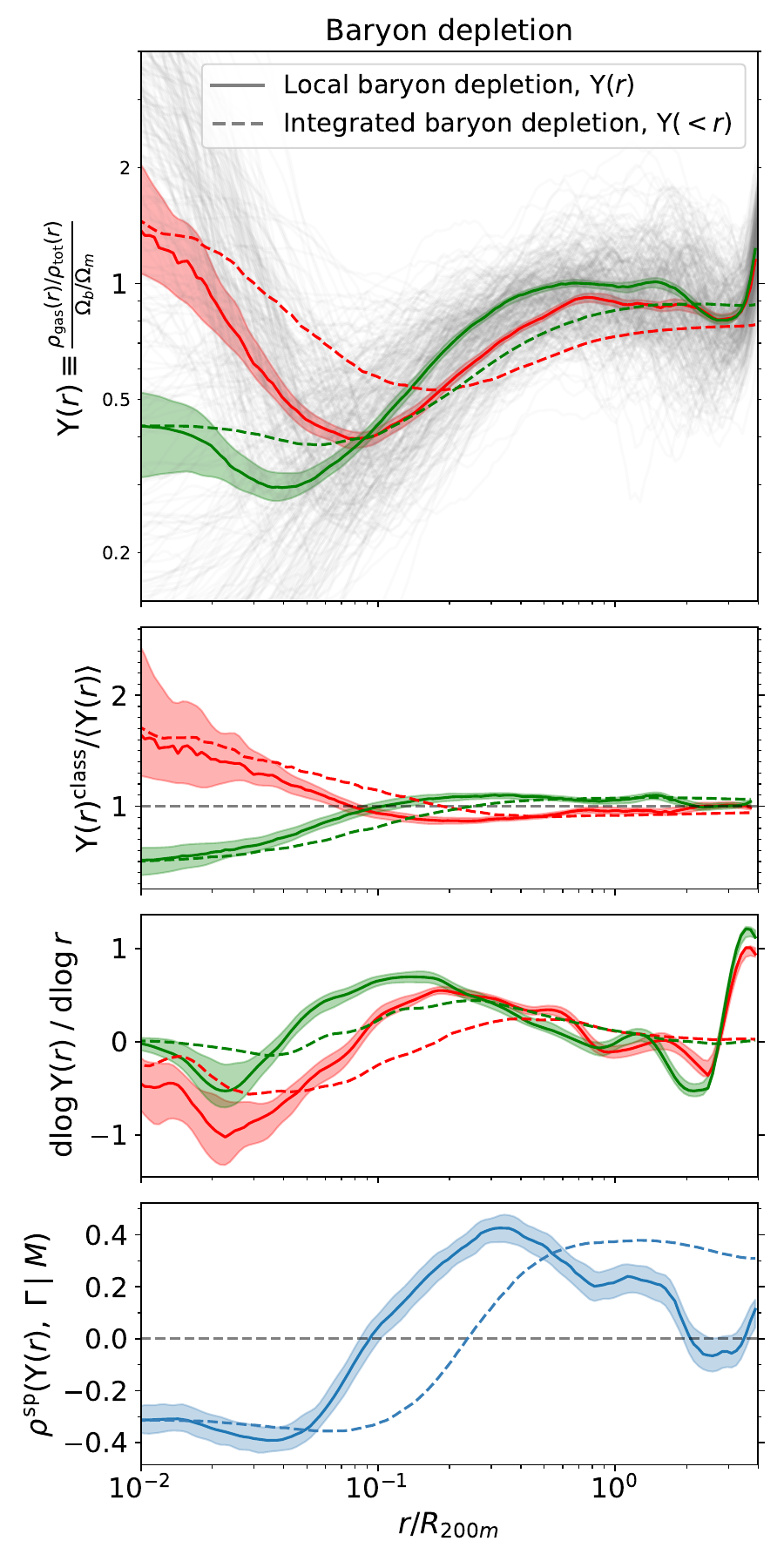}
    \caption{Mass distribution for the highest accreting third (green) and lowest accreting third (red) of the sample. \textit{Left-hand side panel:} gas density. \textit{Central panel:} DM density. \textit{Right-hand side panel:} baryon depletion. Within each column, the top panel presents the profiles stacked over each subsample, together with the whole population (gray lines). The second panel shows each class normalised by the ensemble average. The third panel contains the logarithmic slopes of the profiles, together with an indication of the radius of minimum slope. Last, the bottom panels show the correlation between the profiles and accretion rate $\Gamma_{200m}$ controlling for $M_{200m}$.}
    \label{fig:density_profiles}
\end{figure*}

We start by analysing the effects of cluster assembly, quantified through the accretion rate at an aperture of $R_{200m}$ over the last dynamical time, $\Gamma_{200m}$, as described in Sect. \ref{s:methods.simulation}, on the radial mass distribution of clusters. We consider both, baryonic and DM density profiles, in the left-hand side and central panels of Fig. \ref{fig:density_profiles}, respectively. Within each column, the upper panel displays the individual profiles as gray lines, while red (green) lines with shaded regions correspond to the stacks over the lowest-accreting (highest-accreting) third of the population and their confidence regions.\footnote{In App.~\ref{app:bootstrap} we justify that the effects observed in this and the next section cannot be explained by the mass composition of these two subsamples alone; and hence they point at an actual, dynamical effect.} Subsequent vertical panels in these figures highlight different aspects: the separation between stacked profiles of the two classes (second panel, showing the value of the class-stacked profiles normalised by the stack over the whole sample); the logarithmic derivative of the stacked profiles (third panel); and the partial Spearman rank correlation\footnote{The partial Spearman rank correlation coefficient determines the strength of the correlation between a pair of variables, $x$ and $y$, while controlling for their dependence on a third variable $z$ \citep{Cramer_1946}. See also our \citetalias{Valles-Perez_2025_accr-i} for a more in-depth description.} between the profile values at each $r$ and accretion rate (fourth panel). With respect to this last one, it is worth noting that, throughout this work, we use these partial correlation coefficients to account for the dependences of the profiles on $\Gamma$ that cannot be explained by the mass-dependence of either the profiles (i.e. departures from self-similarity) or accretion rates.

In cluster cores, there is a clear distinctive behaviour between gas and DM. On stacked profiles, low(high)-accreting clusters prefer cuspy (cored) gas density profiles with logarithmic slopes consistent with $\sim 1$ ($\sim 0$) at $\mathcal{R} \equiv r / \mathrm{R_{200m}} \sim 0.01$. This difference is not found on DM profiles, whose asymptotic inner slopes are $\sim - (0.2-0.3)$ regardless of $\Gamma_{200m}$. On a cluster-by-cluster basis, this amounts for a modest (but statistically significant) Spearman rank correlations of the gas central density with accretion rate of $\sim -0.3$, while for DM the correlation is consistent with being null (fourth row of Fig. \ref{fig:density_profiles}). This is a consequence of the ability of intense accretion for offsetting the development of cool cores \citep{Planelles_2009, Hahn_2017, Valdarnini_2021}, while not impacting the central DM distribution; and hence suggesting that the offset must occur through an increased pressure support (contributed by compressive and turbulent heating), rather than just redistribution of the mass by merger-induced advection.

Intermediate regions ($0.1 \lesssim \mathcal{R} \lesssim 1$) are, as often expected \citep{Kravtsov_2012}, more self-similar, thus minimising the relative magnitude of the differences between our two classes. However, looking at the Spearman rank correlations, there are still significant correlations for gas density to be lower at $\mathcal{R} \lesssim 0.2$ and higher at $\mathcal{R} \gtrsim 0.2$ as accretion rate increases, likely as a consequence of gas being displaced to higher radii due to the offset of the cool core. DM presents a different behaviour, its density being anticorrelated with accretion rate up to $\mathcal{R} \lesssim 0.5$, and being especially significant ($|\rho^\mathrm{sp}| \sim 0.5$) within $0.1 \lesssim \mathcal{R} \lesssim 0.3$. 

Interestingly, \citet{CorreaMagnus_2025} recover a similar correlation of gas density with $\Gamma_{200m}$ at intermediate radii to ours; but this correlation also extends to DM in their study, at variance with the results we report here. Their analysis is, however, is based on mean profiles, which are more sensitive to substructure and secondary halos (this being precisely the effect that our median profiles are designed to suppress). We have checked that, if we repeat our analysis with mean profiles, we do recover a positive correlation between $\rho_\mathrm{DM}$ and $\Gamma_{200m}$ at intermediate radii, just as our gas profiles and \citet{CorreaMagnus_2025} DM profiles. This differential behaviour between the mean and median profiles of DM and gas would indicate that gas undergoes a more coherent bulk displacement toward higher radii as a response to intense accretion, whereas the DM profiles mainly reflect the localized influence of the secondary object.

In the outskirts ($\mathcal{R}>1$), the effects of accretion are better highlighted through the logarithmic slopes (third row of Fig. \ref{fig:density_profiles}). In both cases, the radius of minimum (most negative) slope is larger for lower-accreting clusters. In the case of DM profiles, this radius is usually associated with the splashback feature, generated by the accumulation of recently accreted particles in their first apocentric passage \citep{Diemer_2013, Adhikari_2014}. Its anticorrelation with accretion rates has been also highlighted in different studies \citep{Diemer_2014, ONeil_2021}. 

Interestingly, an analogous feature is observed for the gas density profiles, albeit at slightly larger radii. This has been interpreted by several authors as either a response of baryons to the splashback feature of DM, or an indication of the accretion shock \citep{ONeil_2021, Towler_2024}. Unlike \citet{Towler_2024}, we find that this feature does correlate with accretion rate, just as the splashback location does. However, we were unable to reconstruct this feature from hydrostatic equilibrium with the DM density profile alone. On the other hand, its radial location is inconsistent (too internal) to be a signature of the accretion shock. Indeed, the density jump of a strong shock is small ($\lesssim 4$), and could not fully account for the observed effect. Thus, although this feature must emerge as a transition from the one-halo to the two-halo/IGM regimes, its relation to the splashback feature or a shock is not straightforward, and merit a detailed, separate study in the future.

The distinctive behaviour of the two components is reflected in the baryon depletion profile,
\begin{equation}
    \Upsilon(r) = \frac{\rho_\mathrm{gas}(r) / \rho_\mathrm{tot}(r)}{\Omega_b / \Omega_m},
\end{equation}
\noindent presented in the right column of Fig. \ref{fig:density_profiles} for the standard radial profiles (solid lines) and for the cumulative profiles ($\Upsilon(<r)$; dashed lines). Assembly events being the only mechanism capable of offsetting overcooling, dynamically active (relaxed) clusters present a clear underabundance (overabundance) of gas with respect to the cosmic average in the region $\mathcal{R} \lesssim 0.1$.

%----------------------
\subsection{Thermodynamic profiles}
\label{s:results.thermodynamic}
%----------------------

\begin{figure*}
    \centering
    \includegraphics[width=0.33\textwidth]{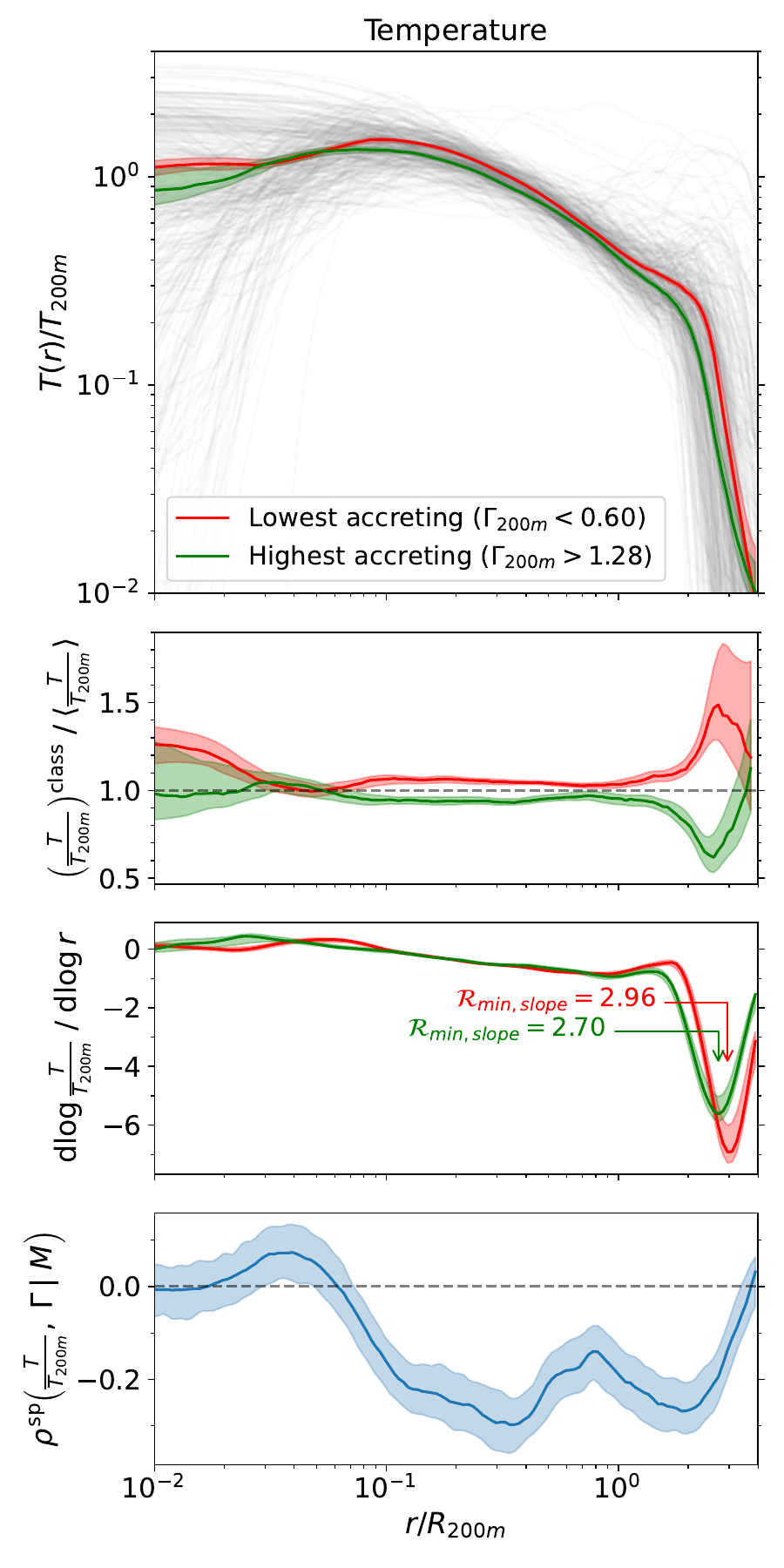}~
    \includegraphics[width=0.33\textwidth]{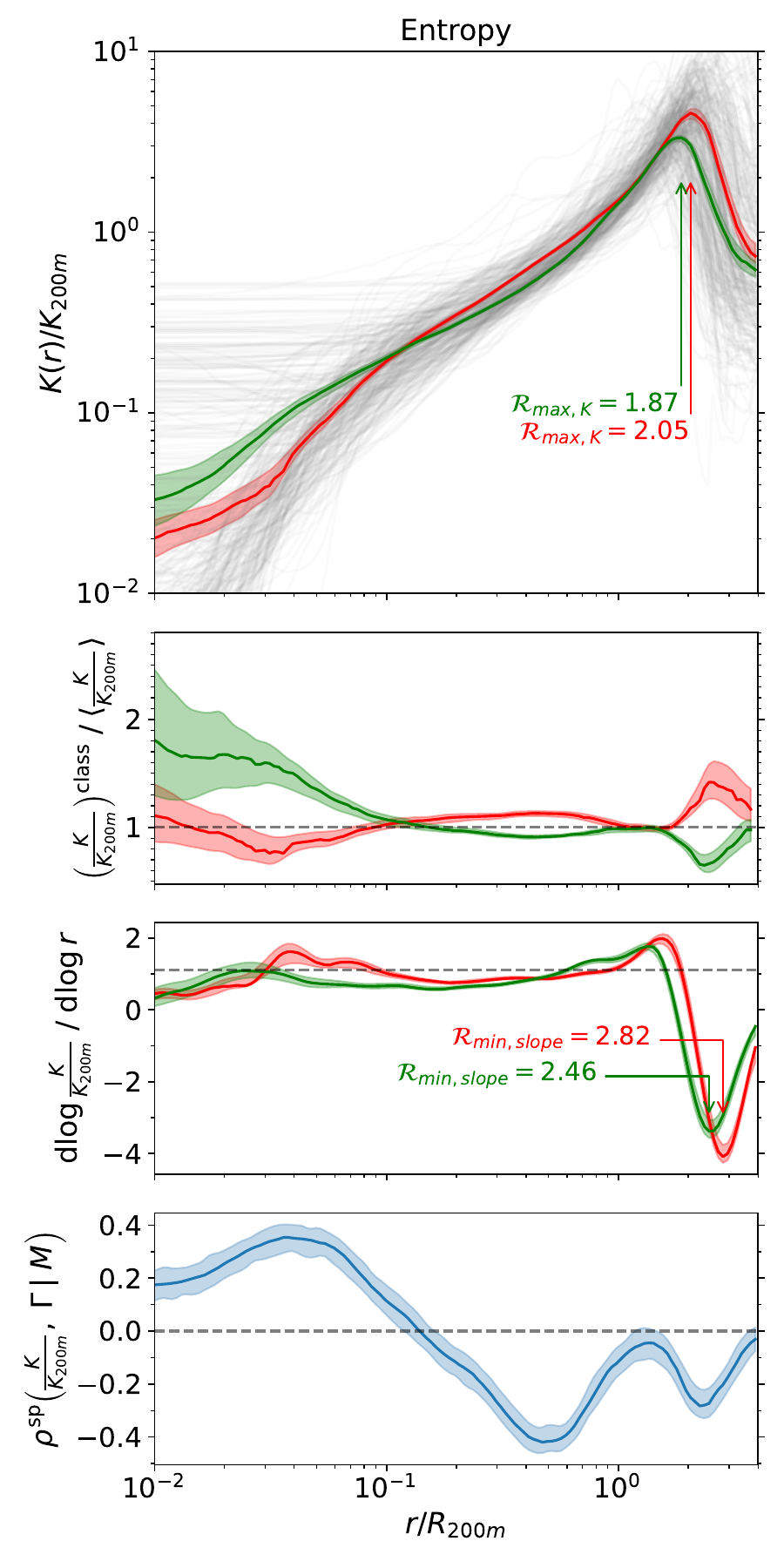}~
    \includegraphics[width=0.33\textwidth]{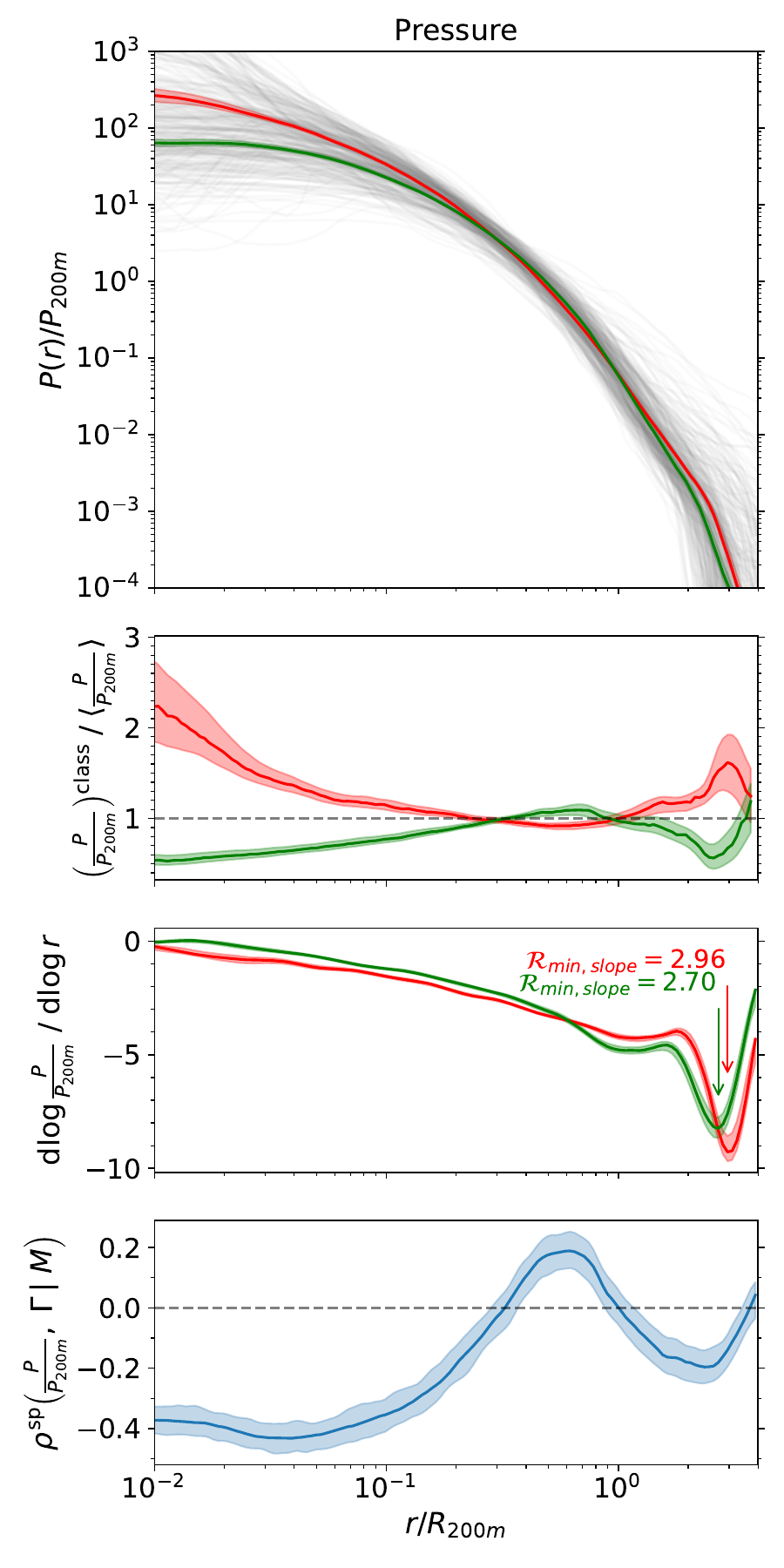}
    \caption{Thermodynamical state of the highest-$\Gamma_{200m}$ third (green) and the lowest-$\Gamma_{200m}$ third (red) of the sample. \textit{Left-hand side panel:} temperature. \textit{Central panel:} entropy. \textit{Right-hand side panel:} thermal pressure. Vertical panels within each column contain the same information as in Fig. \ref{fig:density_profiles}.
    }
    \label{fig:thermo_profiles}
\end{figure*}

We now turn to the analysis of temperature, entropy and pressure profiles, as a summary of the thermodynamical state of the bulk ICM as described by our median profiles. These results are shown in Fig. \ref{fig:thermo_profiles} in an analogous way to Fig. \ref{fig:density_profiles}.

The left-hand side panel presents the temperature profiles for the lowest-accreting (red) and the highest-accreting (green) thirds. Clusters and groups with low accretion rates tend to be hotter --when normalised to their $T_{200m}$-- throughout the whole radial range, with an enhancement of their temperature profiles ranging from $\sim 30\%$ at the core region ($\mathcal{R} \lesssim 0.04$) to $\sim 10\%$ at the intermediate, more self-similar radii. The separation of the logarithmic slopes is fairly limited out to $\mathcal{R} \sim 1$, in such a way that the effect of high accretion on the temperature profiles mainly amounts to a lower normalisation through the virial volume (as confirmed by the moderate, $\rho^\mathrm{sp} \sim -0.3$ anticorrelation). This might impact the integrated temperature and, thus, the $T_X-M$ scaling relation, as also seen for instance by \citet{Chen_2019}, or the $L_X-T_X$ scaling (studied in relation to mergers by \citealp{Planelles_2009}). However, we leave the impact of cluster assembly on the X-ray and SZ scaling relations for a future study.

In the case of entropy profiles (central column of Fig. \ref{fig:thermo_profiles}), instead, we observe a more complex pattern, where the entropy of strongly accreting clusters is consistently higher in the central regions ($\mathcal{R} \lesssim 0.1$), while lower in the rest of the virial volume ($0.1 \lesssim \mathcal{R} \lesssim 1$). The former effect is consistent with the trends observed in the gas density (Sect. \ref{s:results.density}) and temperature profiles and confirms the ability of accretion to disturb cool-cores. The entropy defect at intermediate radii may then be well explained by the merger-driven mixing with the lower-entropy central gas. In this regard, it is also worth discussing the presence of an important ($\sim 25\%$) fraction of entropy cores among the individual profiles. We have verified these to mostly correspond to objects in the high-$\Gamma_{200m}$ subsample, even though also a few low-$\Gamma_{200m}$ objects may display an entropy core. Reasons for this can be mainly due to our definition of accretion rate and aperture choices. For instance, a very tangential merger could increase $\Gamma_{200m}$ while having a very delayed impact on the core; or, conversely, after a frontal merger which disrupts the core significantly, we could expect low (or even negative) accretion rates (e.g., \citealp{Valles-Perez_2020}) if the infalling substructure reaches apocentric distances outside $R_{200m}$.

Pressure profiles (right-most column of Fig. \ref{fig:thermo_profiles}) present their most salient dependence on accretion rate at the cluster centre, where the quotient between the low-$\Gamma$ and the high-$\Gamma$ central pressures reaches a factor of $\sim 4$. In these central regions, this anticorrelation explains almost half of the scatter in the values of $P/P_\mathrm{200m}$, that is, on the departures of the central pressures from self-similarity. This trend is to be expected, since relaxed clusters are almost exclusively pressure-supported in their centres \citep[e.g.][]{XRISM_2025_A2029, XRISM_2025_A2029_b}, while highly-accreting systems tend to have higher central non-thermal pressures \citep{Groth_2025}, as well as departures from hydrostatic equilibrium \citep{Biffi_2016, Angelinelli_2020}. Conversely, there is a slight but significant pressure enhancement in the highly-accreting subsample in the region $0.3 \lesssim \mathcal{R} \lesssim 0.8$, a feature similar to the one observed in the stacked SZ maps of \citet{Monllor_2024}.

Finally, it is instructive to consider the radii of steepest slope of these profiles, often associated with the location of the outermost accretion shock of groups and clusters \citep{Shi_2016, Aung_2021, Anbajagane_2024, Zhang_2025}, even though care must be taken with this identification since the locations change significantly depending on the variable of choice. In general, in all cases (temperature, entropy and pressure) we find a $\sim 10 \%$ decrease in the radial location of this feature for the higher-accreting objects with respect to the lower-accreting ones. A similar effect is observed with the location of the entropy peak, also often associated to the accretion shock radius \citep{Lau_2015, Zhang_2025}, albeit it occurs at smaller radii. While the connection of these features to the accretion shock is far from trivial (even in a spherically averaged profile, it depends not only on the location but also on the underlying slope of the profile), we confirm a clear connection --on an ensemble-averaged sense-- of high accretion rates being able to push inwards by $\sim 10\%$ in radius the baryonic envelope of galaxy clusters and groups.

%----------------------
\subsection{Dependence on individual assembly state indicators}
\label{s:results.indicators}
%----------------------

\begin{figure}
    \centering
    \includegraphics[width=.9\linewidth]{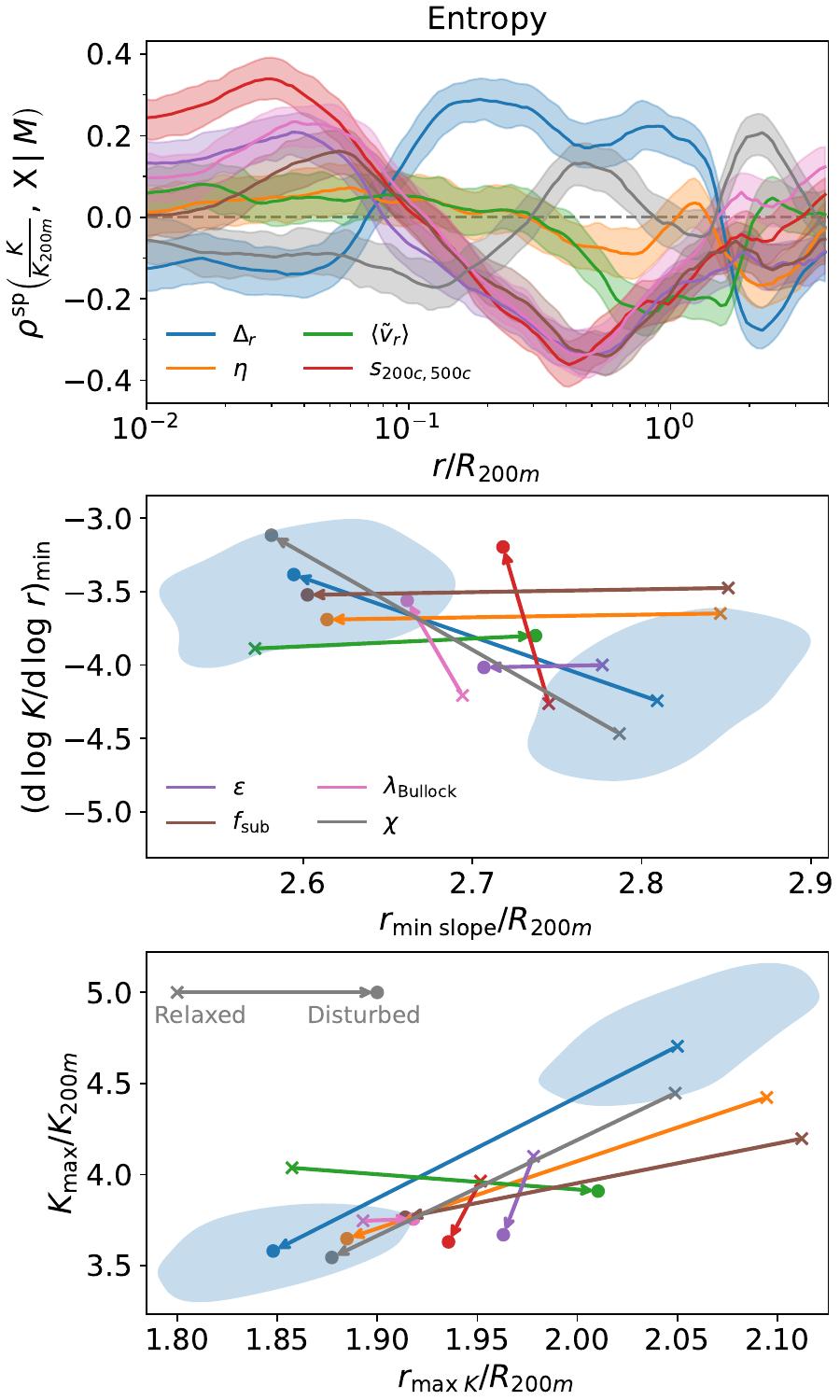}
    \caption{Effect of individual indicators of assembly state on the entropy profiles. \textit{Top panel:} Spearman (partial) correlation coefficients of each indicator and the value of the profiles at each $r/R_{200m}$. Higher magnitudes (either positive or negative) indicate larger influence of the value of the indicator on the profile at this particular radius. \textit{Middle panel:} effect of selecting clusters based on each parameter on the location and the depth of the steepest logarithmic slope of the entropy profile. Crosses correspond to the profile stacked over the one-third most relaxed subsample (according to the given indicator), while filled dots correspond to the most disturbed third. \textit{Bottom panel:} Similar to the middle panel, but with the location and height of the entropy peak. The blue regions in the middle and bottom panel indicate the $68\%$ confidence region for the determination of the corresponding locations over each of the $\Delta_r$-based subsamples (as an example), obtained by bootstrap resampling.}
    \label{fig:entropy_indicators}
\end{figure}

\begin{figure}
    \centering
    \includegraphics[width=.9\linewidth]{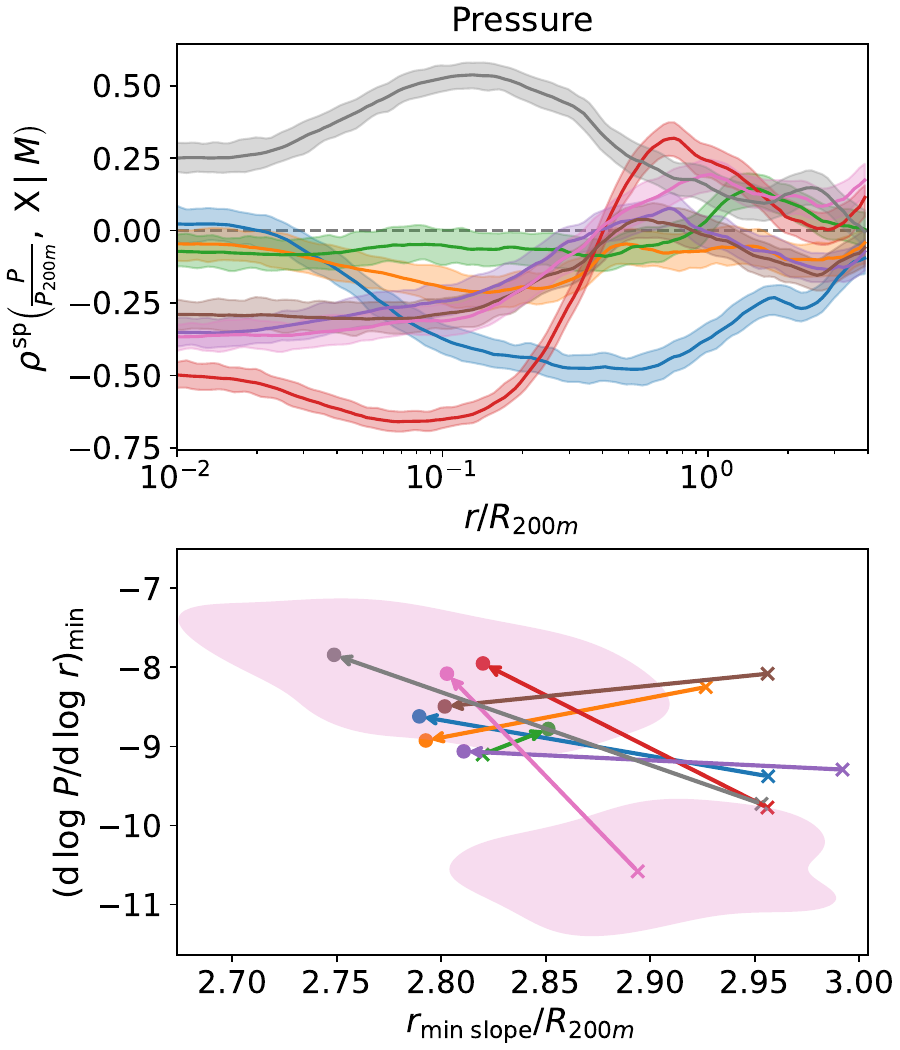}
    \caption{Effect of individual indicators of assembly state on the pressure profiles. \textit{Top panel:} similar to the top panel of Fig. \ref{fig:entropy_indicators}, with the pressure profiles. \textit{Bottom panel:} similar to the middle panel of Fig. \ref{fig:entropy_indicators}, with the pressure profiles. The colour code and other figure elements is kept the same as in the aforementioned figure.}
    \label{fig:pressure_indicators}
\end{figure}

As it has become increasingly evident in the recent literature, most of the quantities commonly used to assess the dynamical state of clusters are only loosely correlated among themselves \citep{Jeeson-Danie_2011, Skibba_2011, Haggar_2020} and reflect different features or moments of the assembly history (\citealp{Wong_2012}; \citetalias{Valles-Perez_2025_accr-i}). Therefore, it is interesting to understand what regions of the radial profile are better constrained by different assembly state indicators. We assess this on the entropy and the pressure profiles in Figs. \ref{fig:entropy_indicators} and \ref{fig:pressure_indicators}, respectively, as two paradigmatic observables in X-ray and SZ. The set of indicators has been extracted from \citetalias{Valles-Perez_2025_accr-i}, where we refer the reader for more details on their definition, and contains the centre offset ($\Delta_r$), virial ratio ($\eta$), mean radial velocity ($\langle \tilde v_r \rangle$), sparsity ($s_{200c,500c}$), ellipticity ($\epsilon$), substructure fraction ($f_\mathrm{sub}$), \citet{Bullock_2001} spin parameter ($\lambda_\mathrm{Bullock}$) and a combined indicator of relaxedness ($\chi$; \citealp{Valles-Perez_2023, Valles-Perez_2025_proceedings}). These quantities are in all cases determined from the three-dimensional description of the DM particle distribution, and taking the virial radius as aperture. As some of them ($\Delta_r$, $\epsilon$) relate to the shape of the cluster, in App. \ref{app:mock-test} we consider a mock scenario to rule out that the dependencies we find here are purely geometric. 

Looking at the entropy profiles, the top panel of Fig. \ref{fig:entropy_indicators} presents the correlation between each of the indicators and the normalised profile $K(r)/K_{200m}$, controlling for the possible mass dependences. This quantity can be roughly interpreted as the fraction of the scatter in $K(r)/K_{200m}$ explained by each indicator. The rather intricate behaviour reflects the broad diversity of information brought up by different indicators. The central regions are better constrained by halo sparsity, with sparser haloes tending to have higher central entropies. At intermediate regions ($0.1 \lesssim \mathcal{R} \lesssim 1$) sparsity is instead anticorrelated with entropy, with halo spin and ellipticity following a very similar radial trend. By contrast, centre offset points in the opposite direction, with higher offset being correlated with higher $K(r)/K_{200m}$ values. In this case, correlation coefficients are not high (reaching only $|\rho^\mathrm{sp}| \sim 0.4$), indicating that, in addition to the effects that can be quantified by these indicators (mostly related to larger-scale assembly), other ones (e.g. radiative cooling) are crucial to set the entropy profile. However, it gives a clear account of the differential behaviour of subsamples built based on different indicators.

The effects of selecting clusters based on these assembly state indicators on the pressure profile (top panel of Fig. \ref{fig:pressure_indicators}) are stronger, especially in the central regions ($\mathcal{R} < 0.2$), where sparsity reaches anticorrelations $|\rho^\mathrm{sp}| \sim 0.5-0.7$, being thus capable of explaining to a considerable extent the variability on the deviations from self-similarity in the profiles. Driven by this indicator, the combination $\chi$ also achieves high ($\rho^\mathrm{sp} \sim 0.5$) correlations in this range. At larger radii $\mathcal{R} \in [0.1, 1]$, also centre offset anticorrelates significantly with the pressure profile, while the effect of other indicators at intermediate and outer radii is limited.

Finally, to study cluster outskirts it is interesting to look at the effect of building samples selected on these parameters on the singular radii studied in Sect. \ref{s:results.thermodynamic}. The second panel of Figs. \ref{fig:entropy_indicators} and \ref{fig:pressure_indicators} show, respectively, how the location and the value of the logarithmic slope minima of the entropy and pressure profiles change from the most-relaxed (crosses) to the most-disturbed (circles) third of the sample according to each indicator. In both cases, the effect of most of the parameters ($\Delta_r$, $\eta$, $f_\mathrm{sub}$, etc.) is pushing the radius of steepest slope inwards, as strongly accreting clusters tend to have slightly less extended atmospheres. However, $\lambda_\mathrm{Bullock}$ and $s_{200c,500c}$ tend to point in the vertical direction, i.e., disturbed objects have a shallower steepening in their outskirts (probably associated to a more aspherical accretion shock shell). 
The third panel of Fig. \ref{fig:entropy_indicators} does the same with the location and value of the entropy peak, which largely mirror the trends seen for the steepest slopes. In all cases, the exception is the mean radial velocity magnitude, which points in the opposite direction, with high $\langle \tilde v_r \rangle$ clusters having more extended atmospheres. This odd behaviour can be attributed to a couple of factors: \textit{(i)} $\langle \tilde v_r \rangle$ is sign-degenerate by construction, meaning that both strong inflows or outflows drive up its value, thus potentially mixing timescales;  \textit{(ii)} as shown by \citetalias{Valles-Perez_2025_accr-i}, it is a very instantaneous indicator, whereas the response of the accretion shock to mergers is a slow process, taking well over a dynamical time.

%----------------------
\subsection{Scatter around the profiles}
\label{s:results.scatter}
%----------------------

So far, we have examined the radial profile trends in connection to cluster assembly. A further key aspect is the scatter around these trends, arising from departures from spherical symmetry.

\begin{figure}
    \centering
    \includegraphics[width=.9\linewidth]{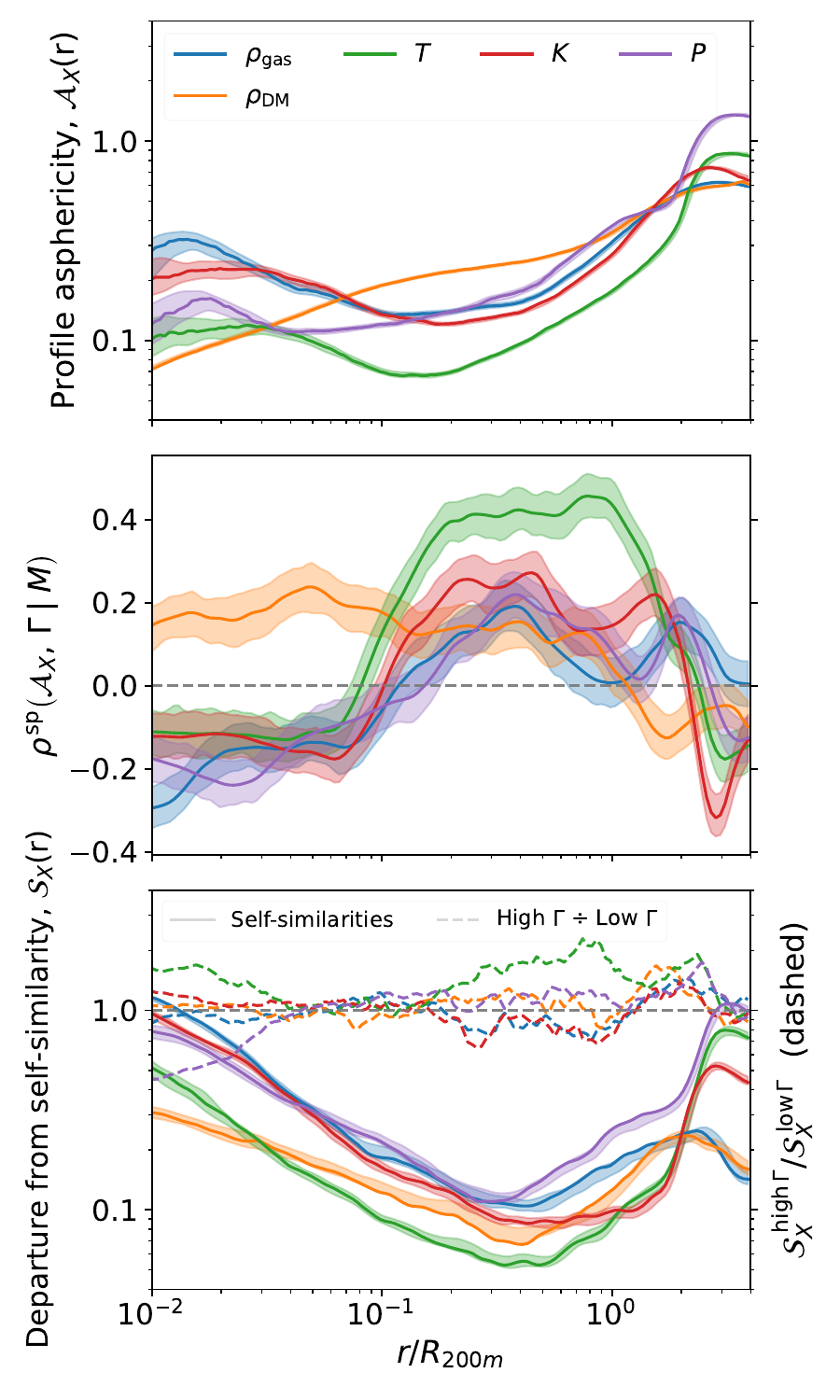}
    \caption{Deviations from sphericity and self-similarity of the clusters' profiles. \textit{Upper panel:} ICM asphericity, $\mathcal{A}(r)$, profiles for different quantities. \textit{Middle panel:} Spearman rank correlation of profile asphericity with accretion rates. \textit{Lower panel:} Departure from self-similarity, $\mathcal{S}(x)$, of the profiles. The dashed lines present the quotient of this quantity among the highest- and lowest-accreting thirds of the sample.}
    \label{fig:scatter}
\end{figure}

In the first panel of Fig. \ref{fig:scatter}, we show the asphericity of the profiles, which we define for a given quantity $X$ as 
\begin{equation}
    \mathcal{A}_X(r) = \frac{1}{2}\log_{10}\frac{X_{84\%}}{X_{16\%}},
    \label{eq:asphericity}
\end{equation}
\noindent with $X_{p\%}$ being the $p$-th percentile of the values of $X$ at radius $r$. That is to say, $\mathcal{A}_X(x)$ is a measure of the logarithmic width of the distribution of $X(\Omega \, | \, r)$, $\Omega$ denoting solid angle. Generally, all gas profiles ($T$, $K$, $P$ and $\rho_\mathrm{gas}$) exhibit similar features, with intermediate radii ($0.1 \lesssim \mathcal{R} \lesssim 1$) being the most spherical and asphericity increasing both towards the centre and the outskirts of the ICM. The latter reflects the increasing anisotropy of the environment, whereas the former likely arises from the complex morphology of cooling flows. These flows are likely to be modified by AGN feedback. Nonetheless, detailed simulations show how gaseous morphology remains complex in the vicinity of AGN \citep{Quilis_2001, Teyssier_2011, Gaspari_2012}. Moreover, although all thermodynamical quantities present broadly similar radial trends, the maximum sphericity (lowest $\mathcal{A}_X(r)$) is achieved at slightly different radii, ranging from $\mathcal{R} \sim 0.05$ for $P(r)$, to $\mathcal{R} \sim 0.2$ for $K(r)$.
Conversely, the DM density anisotropy monotonically increases with radius, thus being most spherical in the centre. This discrepant behaviour is explained by the collisionless nature of DM, where higher densities imply more frequent gravitational collisions (i.e., violent relaxation, collisionless mixing; \citealp{Lynden-Bell_1967, Binney_1987}), and a randomization of particle orbits \citep[e.g.,][]{Navarro_2010}. 

Of interest here is how the asphericity of DM and ICM profiles depends on accretion rates, which we show in the second panel of Fig. \ref{fig:scatter} in a similar way to the bottom panels in Figs. \ref{fig:density_profiles} and \ref{fig:thermo_profiles}. Notably, in the core region, even though correlations are weak ($|\rho^\mathrm{sp}| \sim 0.2$), DM density asphericity correlates with accretion rate, while for gaseous profiles it anticorrelates. This implies that $\sim 20\%$ of the variability in the asphericity of central DM densities is explained by the virial mass increase over the last dynamical time, while clusters that have undergone higher accretion tend to display, somewhat surprisingly, slightly more spherical innermost gaseous distribution. The trend, albeit weak, can be contributed by mergers inducing subsonic turbulence on $\tau_\mathrm{dyn}$ timescales \citep{Vazza_2017, Valles-Perez_2021}, which in turn mixes and homogenizes the baryonic distribution \citep[e.g.][]{Lau_2011}; as well as by the emergence of morphologically-complex cooling flows in very relaxed cluster cores. In the intermediate region, we mildly recover the expected behaviour, where the larger the accretion rates, the more aspherical the ICM gets. This effect is observed both for DM and gas properties, but especially for the temperature profile (which, additionally, tends to be the most spherically-symmetric). The correlations are mild, probably because merging events (which largely determine $\Gamma_{200m}$) tend to impact the tails of the baryonic distributions (beyond the $84\%$ percentile) and may be not captured by our definition in Eq. (\ref{eq:asphericity}) which focuses on the asphericity of the smooth ICM. As a matter of fact, \citet{Vazza_2011} and \citet{Eckert_2012} do recover correlations between dynamical state or cool-coredness (respectively) and scatter around the profiles, likely as a consequence of their definition of the scatter (computed directly from a standard deviation with a suitable masking of outliers), which is more sensitive to extreme values.

Finally, in the third panel of Fig. \ref{fig:scatter} we assess the level of departure from self-similarity of the profiles, which we quantify in a similar way to asphericity, as
\begin{equation}
    \mathcal{S}_X(r) = \frac{1}{2}\log_{10}\frac{\{\mathcal{X}(r)\}_{84\%}}{\{\mathcal{X}(r)\}_{16\%}},
    \label{eq:selfsimilarity}
\end{equation}
\noindent where the notation $\{\mathcal{X}(r)\}_{p\%}$ denotes the $p$-th percentile (over the whole cluster sample) of the values of the median profiles $X(r)$, each normalised by their self-similar value $X_\mathrm{norm}$ (see Sect. \ref{s:methods.stacking}). As widely discussed in the literature \citep[e.g.][]{Ghiradini_2019}, both DM and gaseous profiles are most self-similar at intermediate radii (roughly ${[0.2-0.8] R_{200m}}$ for all quantities considered here). The exception is, perhaps, the entropy profile, whose most self-similar region extends almost up to the accretion shock ($\sim2 R_{200m}$), in line with its most spherically-symmetric region being also located further out. 

In the literature \citep{Pratt_2009, Lovisari_2015}, it is generally observed that relaxed clusters (according to any suitable criterion) behave more self-similarly. To check this, we show, as the dashed lines, the quotient between these departures from self-similarity, $\mathcal{S}_X(r)$, between the highest-accreting third and the lowest-accreting third. These quotients are close to unity, perhaps with the exception of temperature, which is slightly above $1$ throughout the whole radial range. The fact that we do not see broader cluster-to-cluster variance in the median profiles of highly accreting clusters points in the direction of recent literature \citep{Ghiradini_2022, Valles-Perez_2023, Haggar_2024}, in suggesting that there is not a bimodality between disturbed and relaxed clusters, but rather only some --moderate-- continuous trends of the profiles with assembly state.

%----------------------
\subsection{Fits to standard functional forms}
\label{s:results.fits}
%----------------------

We finally shift from the non-parametric profile analyses to the impact of accretion on the parameters of standard fitting formulae.
In order to do so, we have split our complete sample at $z = 0$ into 10 equal-size accretion-rate bins, separated by the deciles in $\Gamma_{200m}$. All fits are performed using a Markov Chain Monte Carlo (MCMC) method. Full details of the fitting procedure and the explicit functional forms are provided in App.~\ref{app:fits}.

\begin{figure}
    \centering
    \includegraphics[width=\linewidth]{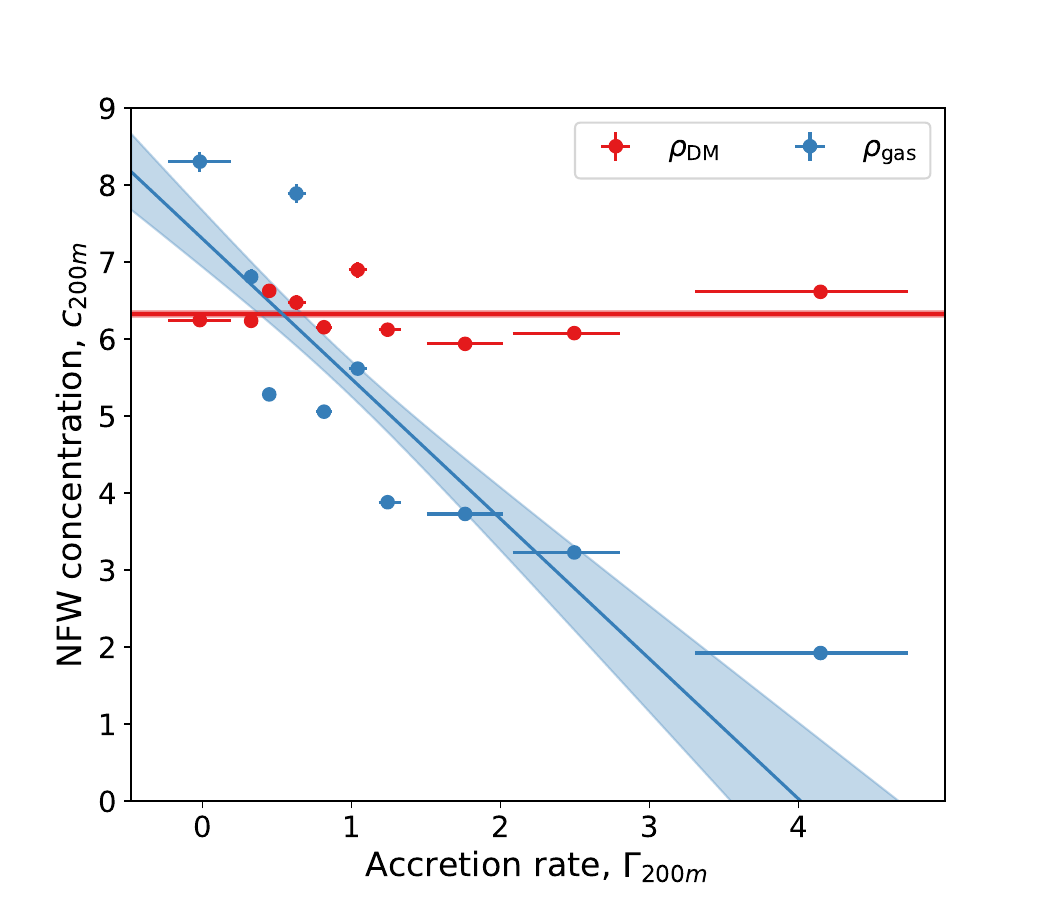}
    \caption{Concentration $c_{200m} = R_{200m} / r_s$, with $r_s$ being the best-fit scale radius of the NFW profile, of the DM (red) and gas (blue) profiles stacked in deciles of accretion rate $\Gamma_{200m}$ over the last dynamical time. Dots indicate the results for each decile of accretion rate, with $1 \sigma$ error bars, while lines are linear fits with their confidence regions.}
    \label{fig:NFW_concentration}
\end{figure}

%-------
\subsubsection{NFW concentration}
\label{s:results.fits.concentration}
%-------

Fig. \ref{fig:NFW_concentration} shows how the concentration $c_{200m}$ of the DM (red) and gas (blue) density profiles changes with the intensity of accretion over the last dynamical time, assuming a \citet[][NFW]{Navarro_1997} functional form (Eq. \ref{eq:NFWform}). For the DM profile, we find no strong trend of $c_{200m}$ with $\Gamma_{200m}$, partly at odds with recent bibliography which suggest that this quantity is driven by major mergers \citep{Wechsler_2002, Wang_2020}. However, this is not surprising, since \citet{Giocoli_2012} showed how $c_{200m}$ depends non-trivially on matter accretion over a combination of timescales. Conversely, for the gas density profiles we do observe a strong anticorrelation\footnote{Even though the NFW is not generally a good fit to the gas density profile, we considered to use them to highlight the different behaviour with respect to the DM profiles. The result obtained here is consistent with the steeper logarithmic slopes of the low-$\Gamma_{200m}$ clusters, as shown in the third panel of Fig. \ref{fig:density_profiles} (left). Incidentally, the NFW form fits remarkably well the cuspy gas profile of low-$\Gamma_{200m}$ clusters.}, in such a way that the larger the accretion rate over the last $\tau_\mathrm{dyn}$, the smaller the concentration parameter (i.e., the larger the scale radius). Both these results are consistent with the findings of Sect. \ref{s:results.density}, where we observed a clear trend for the central gas density profiles (related to the ability of accretion to offset overcooling), while there's no appreciable difference in the DM density slopes.

Overall, we find that DM and gas concentrations in our sample can be well described by:
\begin{gather}
    c_{200m}^\mathrm{DM}(\Gamma_{200m}) = 6.31 \pm 0.09
    , \\
    c_{200m}^\mathrm{gas}(\Gamma_{200m}) = (7.0\pm0.7) - (1.9\pm0.8) \, \Gamma_{200m}
    .
\end{gather}

%-------
\subsubsection{gNFW slopes}
\label{s:results.fits.gNFW}
%-------

\begin{figure*}
    \centering
    \includegraphics[width=\textwidth]{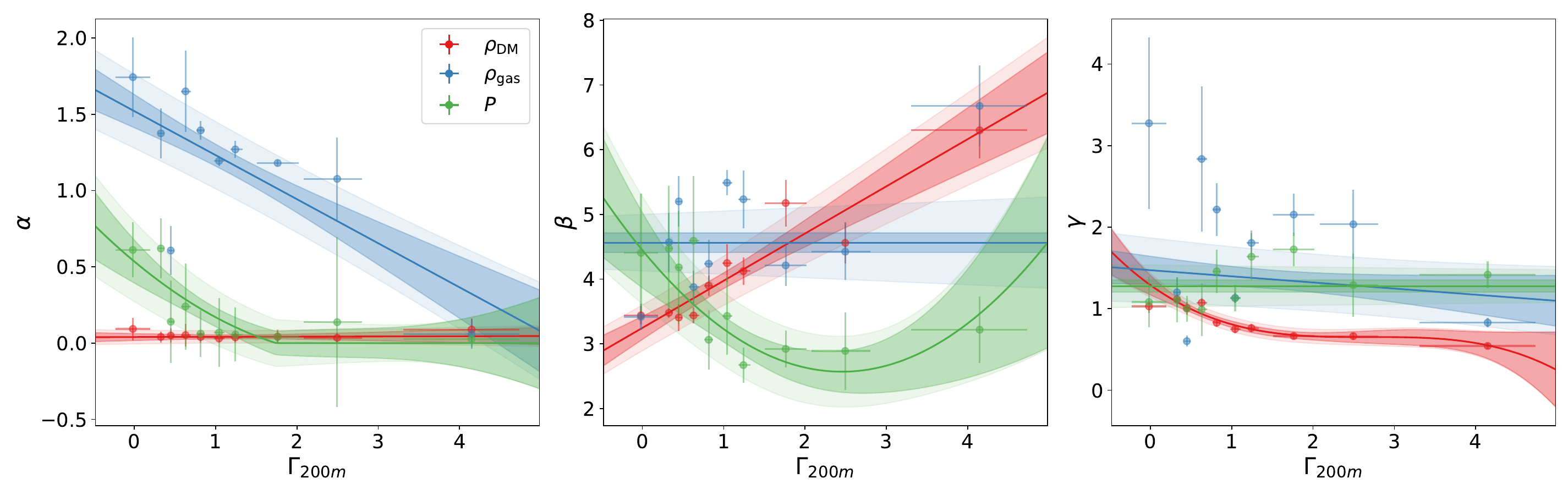}
    \caption{Dependence of the inner slope ($\alpha$, left-hand side panel), the outer slope ($\beta$, middle panel), and the transition parameter ($\gamma$, right-hand side panel) of a gNFW model with accretion rate, for the DM density (red), gas density (blue) and pressure (green) profiles. Dots with error bars represent the results from the fit to the profiles stacked over each decile in $\Gamma_{200m}$. The lines represent least-square polynomial fits, with the dark shaded regions representing the fit uncertainty, and the light shaded regions also contains the uncertainty of the dots added in quadrature (assuming it depends linearly on $\Gamma_{200m}$).}
    \label{fig:gNFW_slopes}
\end{figure*}

Assuming now a generalised NFW form (gNFW, Eq. \ref{eq:gNFWform}), we show in Fig. \ref{fig:gNFW_slopes} the effect of accretion on the internal ($\alpha$) and external ($\beta$) slopes, and the transition sharpness ($\gamma$), for the DM density (red), gas density (blue), as well as the pressure profile (green).

Regarding internal slopes, consistently with the findings of Sect. \ref{s:results.density}, clusters experimenting low accretion develop very steep gas density cusps ($\alpha \sim 2$), which progressively flatten as $\Gamma_{200m}$ increases. The same is not observed for DM profiles, which are consistent with flat cores regardless of accretion rate (possibly partly contributed by mass resolution). Combining this with the central temperature decrease in disturbed clusters, shown in Sect. \ref{s:results.thermodynamic}, pressure profiles exhibit an almost bimodal behaviour, with weak cusps $\alpha \simeq 0.75$ for $\Gamma_{200m} \lesssim 0.5$, and flat cores otherwise. An indication of this was also observed by \citet{Monllor_2024} for the thermal SZ signal.

External slopes of the median profiles (which, as discussed in Sect. \ref{s:results.density}, represent the one-halo term and tend to be steeper than the $\beta=3$ NFW value), present clear dependences with accretion rate. For DM densities, the most relaxed clusters and groups are consistent or only marginally above the NFW value, respectively. However, as accretion rates increase, the outer gNFW slopes get increasingly steeper, even reaching $\beta \gtrsim 6$ for $\Gamma_{200m} \sim 4$. This does not imply that the outer slopes actually reach ${\mathrm{d} \log \rho_\mathrm{DM} / \mathrm{d} \log r = - 6}$ (see Fig. \ref{fig:density_profiles}), as $\beta$ only represents the slope at $r \to \infty$, but emerges as a combination of \textit{(i)} scale radii increase while measured outer slopes (see Fig. \ref{fig:density_profiles}) are kept similar, and \textit{(ii)} splashback radii are smaller, truncating the one-halo term before and thus making the outskirts slightly steeper. Interestingly, the pressure profile outer slope varies in the opposite way, with shallower outer slopes for highly accreting clusters, which again is not observed in the third panel at the right-hand side of Fig. \ref{fig:thermo_profiles}. Overall, these trends warn about the meaning of gNFW parameters, which are highly correlated, when describing the ICM and DM halo structure.

Finally, the parameter controlling the sharpness of the transition ($\gamma$) only presents a noticeable dependence on $\Gamma_{200m}$ for the case of the DM profile, where the transition is smoothed out (lower $\gamma$) for higher-accreting clusters, as a consequence of the loss of sphericity. This trend is much weaker or almost non-existent for the ICM, probably because this medium is already much more spherical due to its collisional nature.

%-------
\subsubsection{Entropy profiles}
\label{s:results.fits.entropy}
%-------

\begin{figure}
    \centering
    \includegraphics[width=0.8\linewidth]{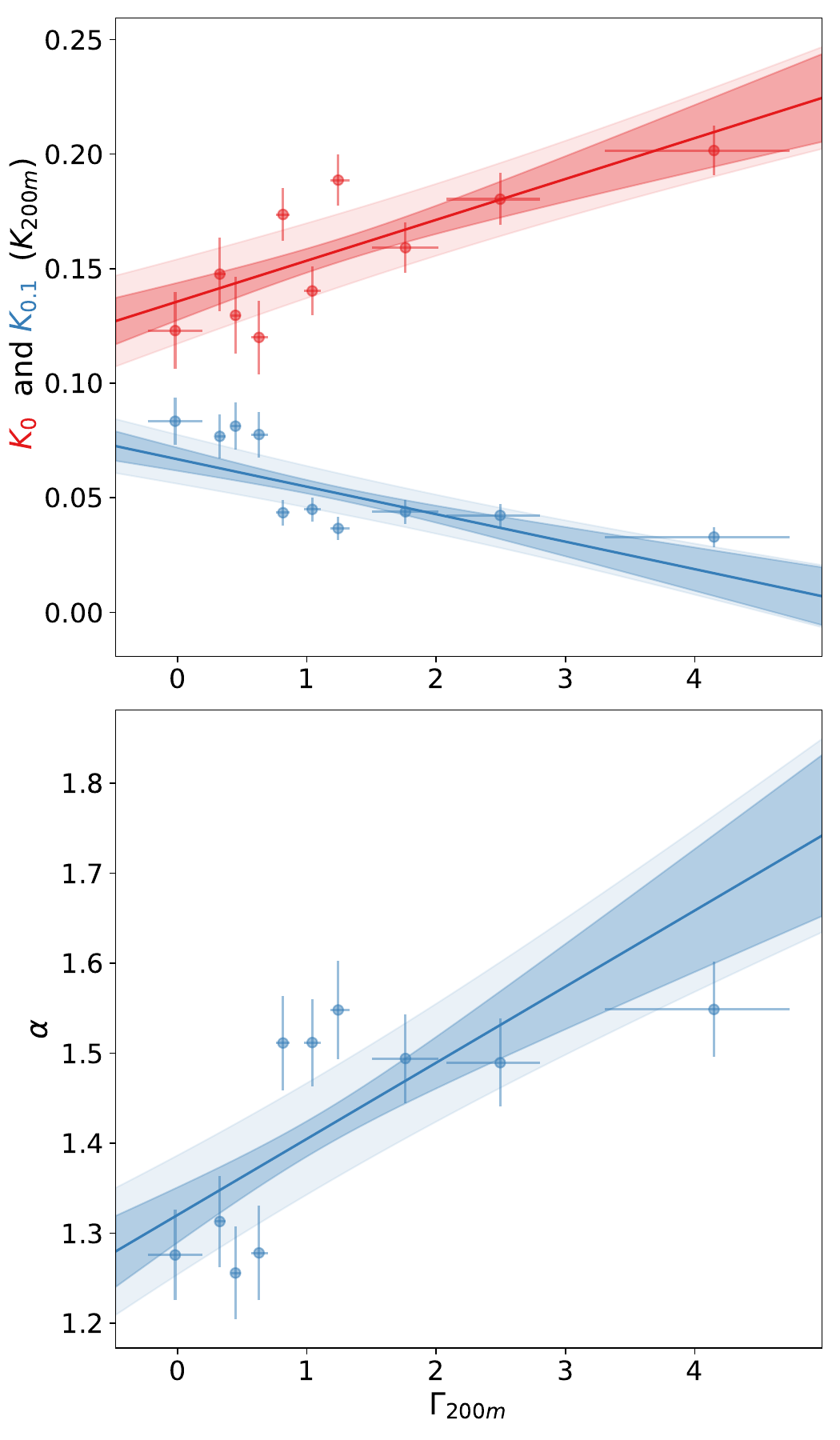}
    \caption{Variation of the core excess entropy ($K_0$; red line in the upper panel), power-law normalisation ($K_{0.1}$, blue line in the upper panel) and power-law index (lower panel) with accretion rate. All plot elements are equivalent to the ones in Fig. \ref{fig:gNFW_slopes}.}
    \label{fig:entropy}
\end{figure}

Lastly, we turn our attention to the entropy profiles, which we fit to the \citet{Donahue_2006} and \citet{Cavagnolo_2009} model (Eq. \ref{eq:cavagnolo}) consisting of a power-law (with normalisation $K_{0.1}$ at $r= 0.1 R_{200m}$ and slope $\alpha$) plus an entropy core ($K_0$). To mitigate the effects of overcooling on the results, we restrict the fits to the interval $r \in [0.1, \,2]R_{200m}$. We show the dependence of $K_0$ and $K_{0.1}$ (red and blue lines, respectively; in units of $K_{200m}$) on $\Gamma_{200m}$ in the upper panel of Fig. \ref{fig:entropy}. 

Entropy floor values present an approximately monotonic increase with $\Gamma_{200m}$, from values of ${K_0 \approx 0.12 K_{200m}}$ to ${K_0 \approx 0.2 K_{200m}}$ when varying from the lowest accreting to the highest accreting deciles. This increase of the core excess, which is a direct manifestation of the role of gas accretion --mainly through mergers-- in offsetting cool cores \citep{Poole_2006, Poole_2008, Planelles_2009, Zinger_2016, Valdarnini_2021}, is compensated by a decreasing normalisation of the power-law range, $K_{0.1}$. Additionally, we also do observe a steepening of the entropy profiles. For relaxed clusters, we find a slope $\alpha \approx 1.25$ (slightly larger than the reference, $\alpha \approx 1.1$, value, \citealp{Voit_2005_entropy}; which is to be expected since cooling times decrease with increasing density, and hence dense gas overcools faster). This slope increases with $\Gamma_{200m}$ reaching values around $\alpha \approx 1.5$ for clusters with $\Gamma_{200m} \gtrsim 1$. Interestingly, these do not seem to follow a continuous trend, but rather a step around $\Gamma_{200m} \sim 1$.

%--------------------------------------------------------------------
\section{Discussion and conclusions}
\label{s:conclusions}
%--------------------------------------------------------------------

Building upon cutting-edge X-ray (\textsc{eROSITA}, \citealp{Merloni_2024}; \textsc{Chex-Mate}, \citealp{Chexmate_2021}) and SZ observational campaigns (e.g. \textsc{Act}, \citealp{ACTDR6}), next-generation observatories (\textsc{NewAthena}, \citealp{Cruise_2025}; \textsc{Simons Observatory}, \citealp{Abitbol_2025}) will provide deeper constraints on ICM thermodynamics, increasing sample size, redshift coverage, and radial apertures. Given the complexity of the physical processing that set the thermodynamical state of the ICM, it is instructive to study in how far the assembly of the cluster itself --set aside internal processes due to galactic physics-- conditions its internal structure. In this work we have aimed at addressing this issue by studying the radial median profiles of DM and ICM properties of a sample of simulated clusters and groups at $z = 0$, spanning masses from $10^{13} \, M_\odot$ to $\sim 7 \times 10^{14} \, M_\odot$, where only cooling --but no feedback processes-- are included besides pure hydrodynamics and gravity. In this way, we can cleanly isolate the effects of gravitational heating, while deferring the assessment of the impact of feedback to a future study. We summarise our main conclusions below, placing them in the framework of recent literature:

\begin{enumerate}
    \item Radial median profiles provide a sound representation of the bulk ICM, largely insensitive to the presence of substructure and clumps without the need of artificial thresholds \citep[e.g.][]{Zhuravleva_2013}, and are insensitive to shot noise independently of binning choices if computed as detailed in Sect. \ref{s:methods.profile}. These profiles largely match the \textit{mode} profiles (which arguably trace the most typical ICM state at a given $r$) while being more robust under larger-scale anisotropy in the outskirts. While the common practice in simulations is to take mean profiles with ad-hoc substructure excision processes, we argue the technique used in this work is also desirable due to its consistency with the azimuthal median technique in observational data \citep{Eckert_2015}.
    \item Accretion affects DM and gas density profiles differently. Central gas reduced by a factor of up to 4 between the lowest and highest accreting subsamples, effectively transforming strong cusps into flat cores; while DM central densities are not significantly affected by accretion. This partly contrasts with \citet{Lau_2015}, who reported inner gas density slopes remaining unaffected by accretion (their figure 8b), likely caused by their non-radiative set-up. At intermediate radii ($0.1 \lesssim r/R_{200m} \lesssim 1$) there's a tendency for highly accreting clusters to have their gas content displaced to higher radii. 
    \item Central temperatures do not strongly vary with accretion rate, but its normalisation at intermediate radii decreases by $\sim 10\%$ for the highly-accreting subsample. Entropy profiles, in turn, do exhibit a clear central increase for highly-accreting clusters, as the central gas mixes more effectively with the higher-entropy surroundings. Consistently, at intermediate radii, there is a $\sim (10-20)\%$ decrease in the entropy of disturbed clusters. Lastly, pressure profiles reveal their largest dependence on accretion in the centre, up to a factor of $\sim 4$ between the two extreme classes (as it was also found for thermal SZ profiles, for instance, by \citealp{Monllor_2024}). In general, these trends were not captured by the analyses of \citet{Lau_2015}, most likely because the absence of cooling prevents the buildup of dense and cold central gas, that accretion would otherwise reheat.
    \item In the outskirts, high accretion shifts the radii of steepest slope (both for $\rho_\mathrm{DM}$ and for $\rho_\mathrm{gas}$) inward. This result had already been found for the $\rho_\mathrm{DM}$, where this steepest slope roughly corresponds to the splashback feature \citep[e.g.][]{More_2015, Shin_2023}. For the thermodynamical profiles (temperature, entropy and pressure profiles), the radius of steepest slope responds to the average location of the accretion shock. It does shrink --albeit to a smaller extent to $R_\mathrm{sp}$, $\sim 10\%$-- with increasing accretion rate, as expected from theoretical considerations \citep{Shi_2016} and also reported by \citet{Aung_2021} and \citet{Zhang_2025}.
    \item Selection of clusters based on indicators of assembly state bears a non-trivial effect on the profiles (Figs. \ref{fig:entropy_indicators} and \ref{fig:pressure_indicators}). For instance, while entropy in the $0.1 R_{200m} \lesssim r \lesssim R_{200m}$ range moderately anticorrelates with sparsity, it is positively correlated with centre offset, despite both $s_{200c,500c}$ and $\Delta_r$ being a measure of dynamical unrelaxedness. Moreover, the correlations do not tend to be strong (in line with, e.g., \citealp{Lau_2021}, their figure 4).
    
    These quantities, insofar as they are proxies for the halo and ICM mass growth, carry an imprint on the characteristic outermost radii delimiting the extent of the ICM (entropy peak, steepest slopes). In general, less relaxed clusters (according to any indicator) exhibit baryonic boundaries smaller by up to $\sim 10\%$. However, other quantities (e.g. $\Delta_r$ or $\lambda_\mathrm{Bullock}$) are more indicative of the sharpness of the features, suggesting that they are predictive of the sphericity of the accretion shock.
    \item Departures form sphericity in the baryonic profiles are minimum at intermediate ($0.1 R_{200m} \lesssim r \lesssim R_{200m}$) radii --where the profiles are also more self-similar-- and increase both towards the core and the outskirts due to complex cooling flow morphologies and environment anisotropy, respectively. In contrast, DM density is most spherical at the centre. Correlations of asphericity with accretion rates over the last dynamical time are rather weak, as a result of a combination of physical effects and our definition of asphericity, perhaps with the clearest signal on the temperature profile at intermediate radii. Looking at the intrinsic scatter over the cluster population, we recover the trends observed by \citet{Ghiradini_2019} on the X-COP sample, namely a log-parabolic trend for the scatter, with temperature and pressure profiles behaving, respectively, as the most and least self-similar quantity.
    \item When profiles are stacked on deciles of $\Gamma_{200m}$, we see clear signatures of accretion on the estimation of parameters from widely used functional forms. In particular, the concentration of a NFW model fitted to gas density reveals a tight dependence of $c_{200m}$ on $\Gamma_{200m}$, while the same does not happen for the DM density profiles. This might be the case because the value of concentration of the gaseous profile depends on the emergence of a cusp, driven by (over)cooling, while in the case of DM concentration, it has been reported how it depends non-trivially on the combination of several timescales \citep{Giocoli_2012}.

    When leaving the slopes as free parameters (gNFW), we recover a tendency for gas to develop strong cusps ($\alpha \simeq 1.5$) in the absence of accretion, that are progressively transformed into cores for high accretion. External slopes, especially in the case DM density, exhibit a strong dependence on accretion rates, with sharper truncations for higher-accreting obejcts. Entropy profiles adhere to the \citet{Cavagnolo_2009} form only within $0.1 R_{200m} \lesssim r \lesssim 2 R_{200m}$, with a clear indication of the offset of cool cores by strong accretion. 
\end{enumerate}

Several limitations of the present analysis should be kept in mind. While the simulation set-up employed here is ideal to isolate gravitational heating on the ICM, future work will need to incorporate its complex interplay with stellar and AGN feedback at the group and cluster scales, which are key to obtaining ICM distributions consistent with observations \citep{McNamara_2012, Fabian_2012, Planelles_2014}, especially the innermost regions. A second caveat concerns our assumption of self-similar scaling to stack the profiles. A number of recent works \citep{Pratt_2010, Riva_2024} have highlighted how substantial deviations from the theoretical \citep{Kaiser_1986} self-similar scaling with mass result in a decreased intrinsic scatter around the profiles. Though we do not address this here due to space and scope limitations, this remains a key issue, and represents an important direction for future work. Finally, a remaining issue of high interest is assessing to what extend the trends that we have identified in three-dimensional profiles remain in projection. Advancing along these lines will be essential in order to extract the maximum information from current and future observational surveys of the thermal ICM.

%--------------------------------------------------------------------
\begin{acknowledgements}
We are grateful to Annalisa Bonafede and to the anonymous reviewer for their careful reading of this manuscript and valuable suggestions, and to Franco Vazza and Marco Balboni for insightful discussions about this work. This work has been supported by the Agencia Estatal de Investigación Española (AEI; grant PID2022-138855NB-C33), by the Ministerio de Ciencia e Innovación (MCIN) within the Plan de Recuperación, Transformación y Resiliencia del Gobierno de España through the project ASFAE/2022/001, with funding from European Union NextGenerationEU (PRTR-C17.I1), and by the Generalitat Valenciana (grant CIPROM/2022/49). DVP acknowledges partial support from Universitat de València through an Atracció de Talent fellowship. The simulation and part of the analysis have been carried out using the supercomputer Lluís Vives at the Servei d'Informàtica of the Universitat de València.

This research has been made possible by the following open-source projects: Numpy \citep{Numpy}, Scipy \citep{Scipy}, Matplotlib \citep{Matplotlib}, FFTW \citep{FFTW}, \texttt{emcee} \citep{Foreman-Mackey_2013}.
\end{acknowledgements}
%--------------------------------------------------------------------
\bibliographystyle{aa}
\bibliography{aa57496-25}

\clearpage
\appendix

%--------------------------------------------------------------------
\section{Self-similar normalisation constants}
\label{app:self-similar}
%--------------------------------------------------------------------

All density profiles in this work have been normalised by the background density of the corresponding component (i.e., $f_b \rho_B$ for gas density; or $f_\mathrm{DM} \rho_B$ for DM density profiles\footnote{$f_\mathrm{DM} = 1 - f_b$ is the DM mass fraction.}) in order to stack them or compute correlation coefficients. 

Results regarding temperature, entropy and pressure profiles are also normalised by their corresponding self-similar values at the chosen overdensity of $\Delta_m = 200$, which we give below for completeness.

In these equations, $G$ and $k_B$ are the gravitational and Boltzmann constants, $m_p$ the proton mass, $\mu$ the mean molecular weight of a fully ionised plasma, $n = \rho_\mathrm{gas} / (\mu m_p)$ the particle number density and $n_{\Delta_m}$ its value corresponding to a mean matter overdensity of $\Delta_m$.

\begin{strip}
\rule{0.9\textwidth}{0.4pt}
\begin{equation}
    k_B T_{\Delta_m} = \frac{G M_{\Delta_m} (\mu m_p)}{2 R_{\Delta_m}} =
    0.93211 \, \mathrm{keV} \, 
    \left( \frac{\Delta_m}{200} \right)^{1/3} 
    \left( \frac{h}{0.678} \right)^{2/3} 
    \left( \frac{\mu}{0.6} \right) 
    \, (1+z) \, 
    \left( \frac{M_{\Delta_m}}{10^{14} M_\odot} \right)^{2/3}
    \label{eq:T200m}
\end{equation}
\begin{equation}
    K_{\Delta_m} = k_B T_{\Delta_m} n_{\Delta_m}^{-2/3} = 
    491.127 \, \mathrm{keV \, cm^2} \, 
    \left( \frac{\Delta_m}{200} \right)^{-1/3} 
    \left( \frac{f_b}{0.155} \right)^{-2/3} 
    \left( \frac{\Omega_{m}(z=0)}{0.31} \right)^{-2/3} 
    \left( \frac{h}{0.678} \right)^{-2/3} 
    \left( \frac{\mu}{0.6} \right)^{5/3} 
    \, (1+z)^{-1} \, 
    \left( \frac{M_{\Delta_m}}{10^{14} M_\odot} \right)^{2/3}
    \label{eq:K200m}
\end{equation}
\begin{equation}
    P_{\Delta_m} = n_{\Delta_m} k_B T_{\Delta_m} = 
    7.70685 \times 10^{-5} \, \mathrm{keV \, cm^{-3}} \, 
    \left( \frac{\Delta_m}{200} \right)^{4/3} 
    \left( \frac{f_b}{0.155} \right) 
    \left( \frac{\Omega_{m}(z=0)}{0.31} \right)
    \left( \frac{h}{0.678} \right)^{8/3} 
    \, (1+z)^{4} \, 
    \left( \frac{M_{\Delta_m}}{10^{14} M_\odot} \right)^{2/3}
    \label{eq:P200m}
\end{equation}
\rightline{\rule{0.9\textwidth}{0.4pt}}
\end{strip}

%--------------------------------------------------------------------
\section{Mass-matched bootstrap null test}
\label{app:bootstrap}
%--------------------------------------------------------------------

In Sects. \ref{s:results.density} and \ref{s:results.thermodynamic}, we examined how the density and thermodynamical profiles (respectively) differ among the highest- and the lowest-accreting cluster subsamples. Selecting clusters by accretion rate could introduce some mass dependence, since most massive clusters at $z \simeq 0$ are often the outcome of a recent major merger. In turn, as real clusters deviate from self-similarity, the mass dependence among the subsamples could have an influence on the stacked profiles. It is therefore important to rule out that the reported effects are due to the mass composition of the samples.

To address this issue, we perform a null test by constructing control samples, matched to the $M_{200m}$ distribution of the low-$\Gamma_{200m}$ and the high-$\Gamma_{200m}$ samples, but selected without reference to $\Gamma_{200m}$. Any residual difference in the profiles between these control samples would then indicate a contribution from their different mass compositions, rather than from $\Gamma_{200m}$ itself.

\begin{figure*}
    \centering
    {\includegraphics[width=0.3\textwidth]{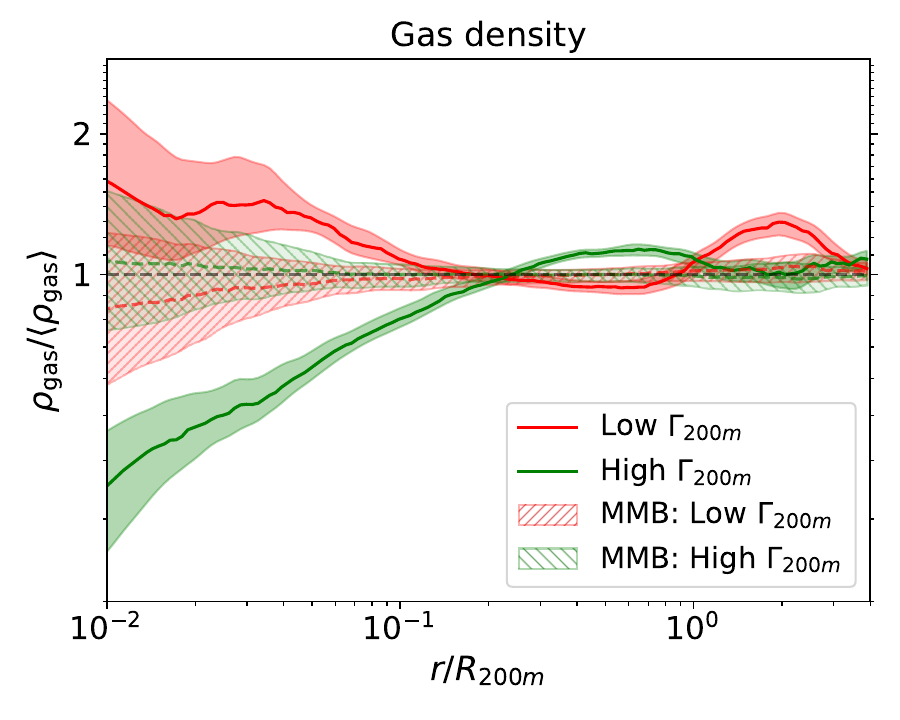}~
    \includegraphics[width=0.3\textwidth]{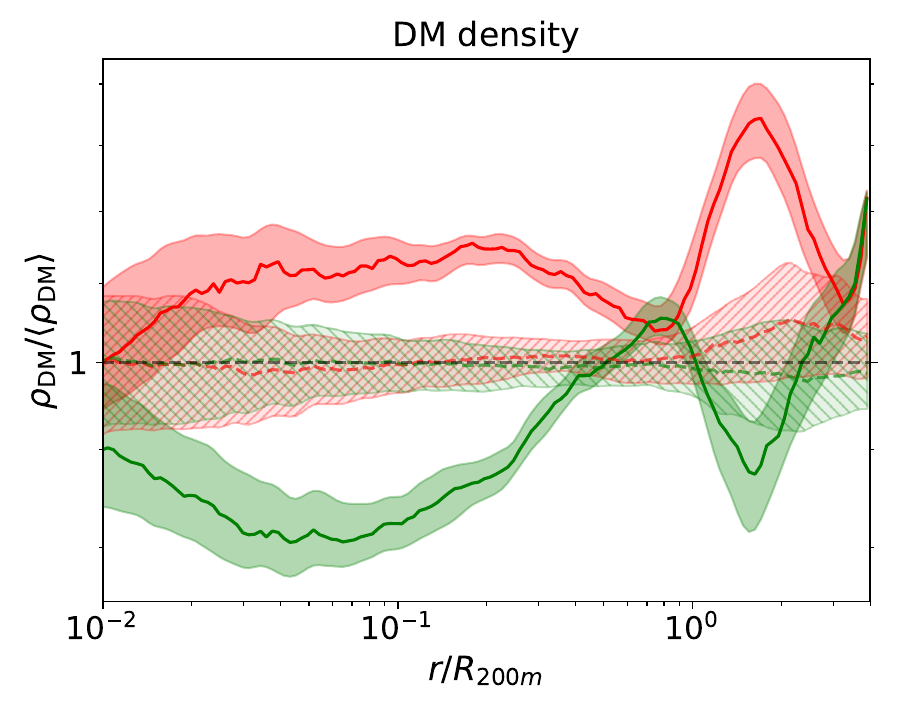}~
    \includegraphics[width=0.3\textwidth]{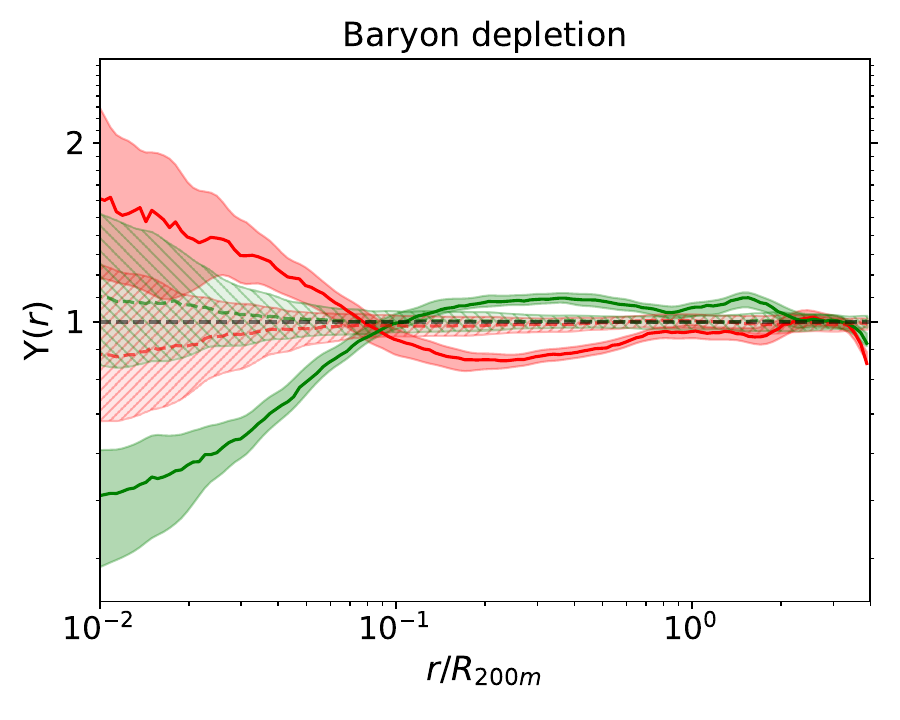}}
    {\includegraphics[width=0.3\textwidth]{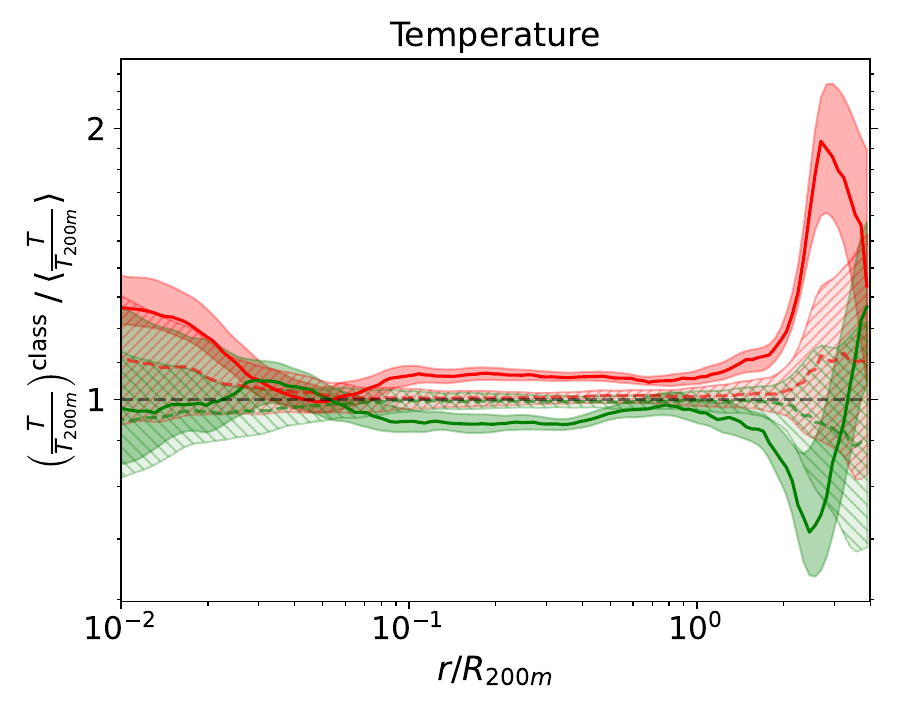}~
    \includegraphics[width=0.3\textwidth]{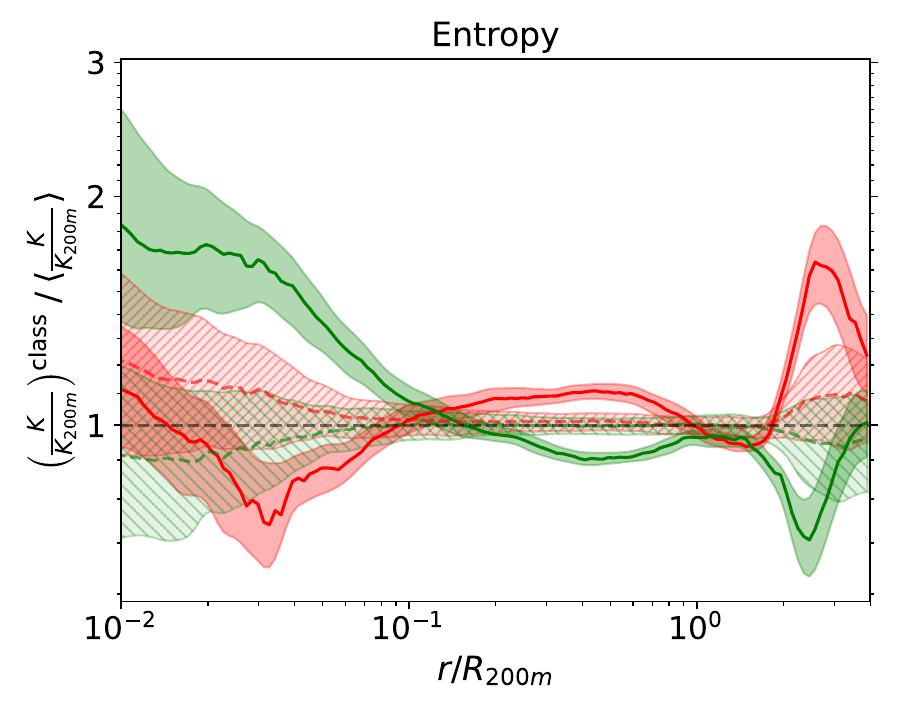}~
    \includegraphics[width=0.3\textwidth]{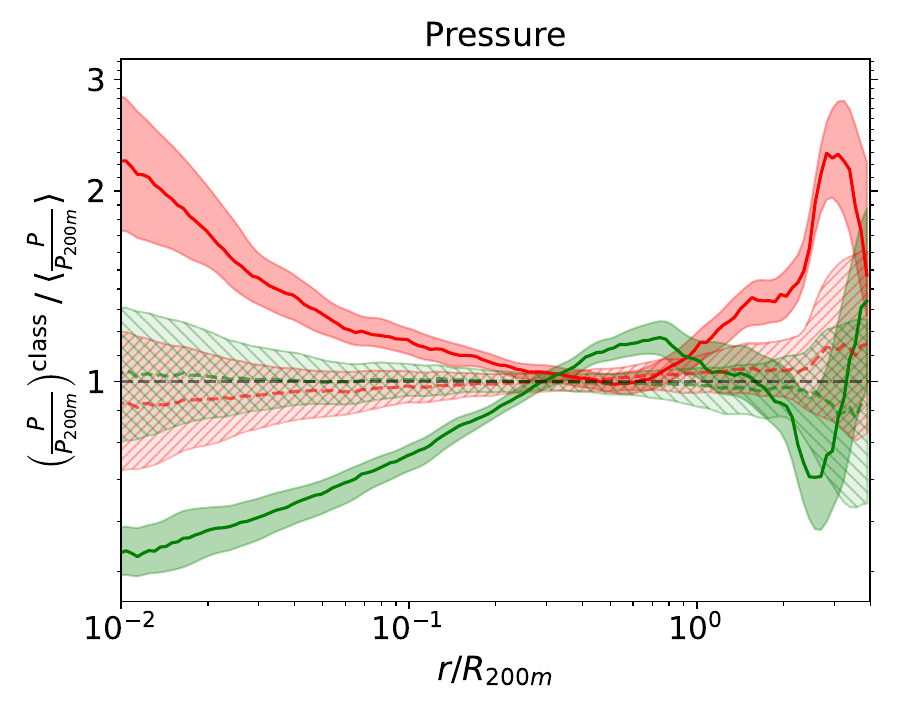}}
    \caption{Mass-matched null bootstrap test. Each panel here is equivalent to the second vertical panel of each of the columns in Figs. \ref{fig:density_profiles} and \ref{fig:thermo_profiles}. Control samples matched in $M_{200m}$ (hatched regions for the $(16-84)\%$ percentiles, dashed lines for the medians) show the expected distributions under the null hypothesis of no $\Gamma_{200m}$ dependence. Solid lines with shaded regions reproduce the original results (red: low-$\Gamma_{200m}$, green: high-$\Gamma_{200m}$). The presence of a separation of the solid lines, beyond the hatched regions, proves that the observed effects cannot be explained by just the mass distribution of the low- and high-$\Gamma_{200m}$ subsamples.}
    \label{fig:bootstrap_results}
\end{figure*}

We repeat the procedure above for building the control samples $N_\mathrm{boots} = 1000$ times, so as to obtain the distribution of the null hypothesis (i.e., no effect of $\Gamma_{200m}$ on the profiles). These distributions are shown in Fig. \ref{fig:bootstrap_results}, which are equivalent to the second vertical panels of Figs. \ref{fig:density_profiles} and \ref{fig:thermo_profiles}. The results for the null hypothesis are shown as the hatched regions, which enclose the $(16-84)\%$ percentiles, with the dashed coloured lines indicating the median, and following the colour coding of the original figures (red for low-$\Gamma_{200m}$, green for high-$\Gamma_{200m}$). The solid lines with solid shaded regions are the original results shown in the aforementioned figures of the main text.

Thus, our results will be able to confirm a genuine signal of the selection of clusters based on recent accretion, if the actual segregation of the profiles (solid lines/regions) is not compatible with the effect of mass selection (hatched regions and dashed lines). 

The effect of the different mass composition of the two subsamples on the profiles is, in all cases, of very limited significance. Regarding the gas density profiles, the only noticeable trend (which however falls within the sample variance, i.e., they are not significant) is an increase in the central gas density for the null sample matched to the high-$\Gamma_{200m}$ distribution, which is the contrary trend to what we have found to occur for the actual samples. A similar effect happens for the $\Upsilon(r)$, $K(r)$ and $P(r)$ profiles, implying that the patterns that we have reported in the main text cannot be attributed to the mass composition of the samples.

At intermediate radii, the mass-matched control samples have virtually identical profiles for all the studied quantities, which is to be expected since in these regions the profiles are the most self-similar. 

Finally, in the cluster outskirts, there is a small effect of the mass selection of the null subsamples on the DM density, temperature, entropy and pressure profiles, that matches what we observed on the actual subsamples (i.e., we see an increase in these quantities, approximately at the locations of the splashback and the shock radii, for the null sample matched to the low-$\Gamma_{200m}$ mass distribution). Again, these differences are not significant given their confidence intervals.  In contrast, in all four cases, the actual magnitude of the effect of selecting by $\Gamma_{200m}$ overshoots the signal that could be solely attributed to mass, and these actual effects are statistically significant. Thus, while a small part of the signal could be attributed to the mass composition, the test shows that, also in this case, our results in the main text reflect the dynamical effect of recent accretion, rather than just a selection bias.

%--------------------------------------------------------------------
\section{Geometric effects of centre offset and ellipticity on the density and entropy median profiles}
\label{app:mock-test}
%--------------------------------------------------------------------

In Sects. \ref{s:results.density} and \ref{s:results.thermodynamic}, we have studied the effects of gas accretion over the last dynamical time on the density and entropy profiles, among others. Subsequently, in Sect. \ref{s:results.indicators} we have explored the effect of selecting clusters based on several assembly state indicators on entropy and pressure profiles. Among these indicators, some of them contain geometric information about the ICM (i.e., the ellipticity $\epsilon$; or the centre offset $\Delta_r$). It is thus interesting to explore to what extent the effects that we find are explicitly caused by geometry (i.e. we are computing a radial profile of a system that progressively loses its spherical symmetry) or if there is an actual dynamical effect.

To do so, we have realised a spherically-symmetric gNFW density profile with inner slope $\alpha=0.5$, concentration $c_\mathrm{vir} = 3$, and total mass $M_\mathrm{vir} = 4 \times 10^{14} \, M_\odot$; and an entropy profile following Eq. (\ref{eq:cavagnolo}) with $K_0 = 40 \, \mathrm{keV \, cm^2}$, $K_{0.1}= 50 \, \mathrm{keV \, cm^2}$ and $\alpha=1.1$. The entropy profile is then interrupted at $\sim 2.5 R_\mathrm{vir}$ with a sigmoid function to mock the accretion shock. Subsequently, we have computed radial profiles following the methodology described in Sect. \ref{s:methods.profile}, by considering the following geometric transformations:

\begin{itemize}
    \item A centre shift, $\Delta_r$, which is realised by computing the profiles considering as the centre a position displaced $\Delta_r R_\mathrm{vir}$ from the mock centre (density peak).
    \item An ellipticity, $\epsilon$, achieved by rescaling the $x$ and $z$ coordinates as $x' = x \sqrt{1 + \epsilon}$, $z' = z / \sqrt{1+\epsilon}$ before evaluating the value of the radial profile at each point.
\end{itemize}

\begin{figure*}
    \centering
    \includegraphics[width=0.4\linewidth]{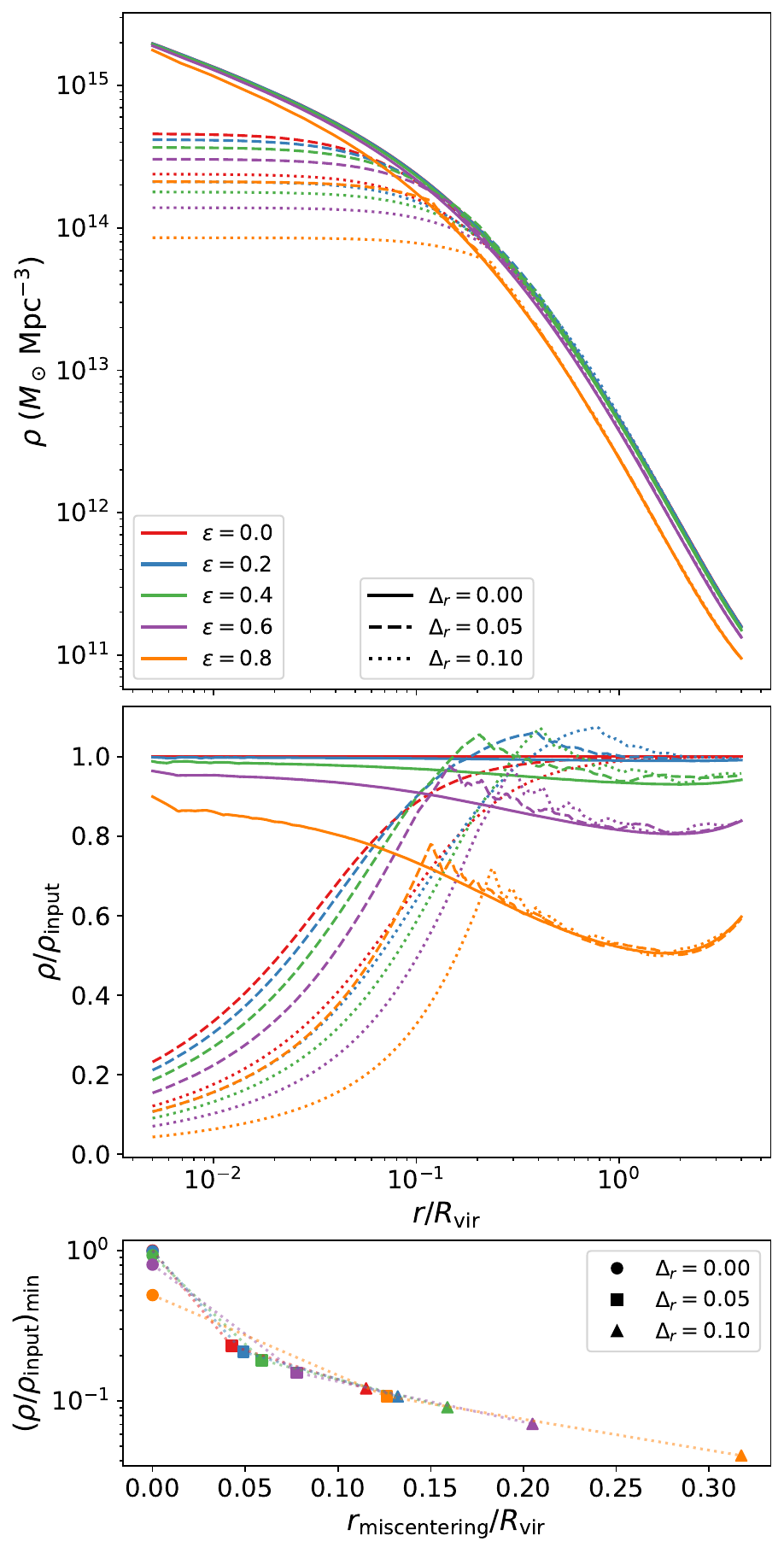}~
    \includegraphics[width=0.4\linewidth]{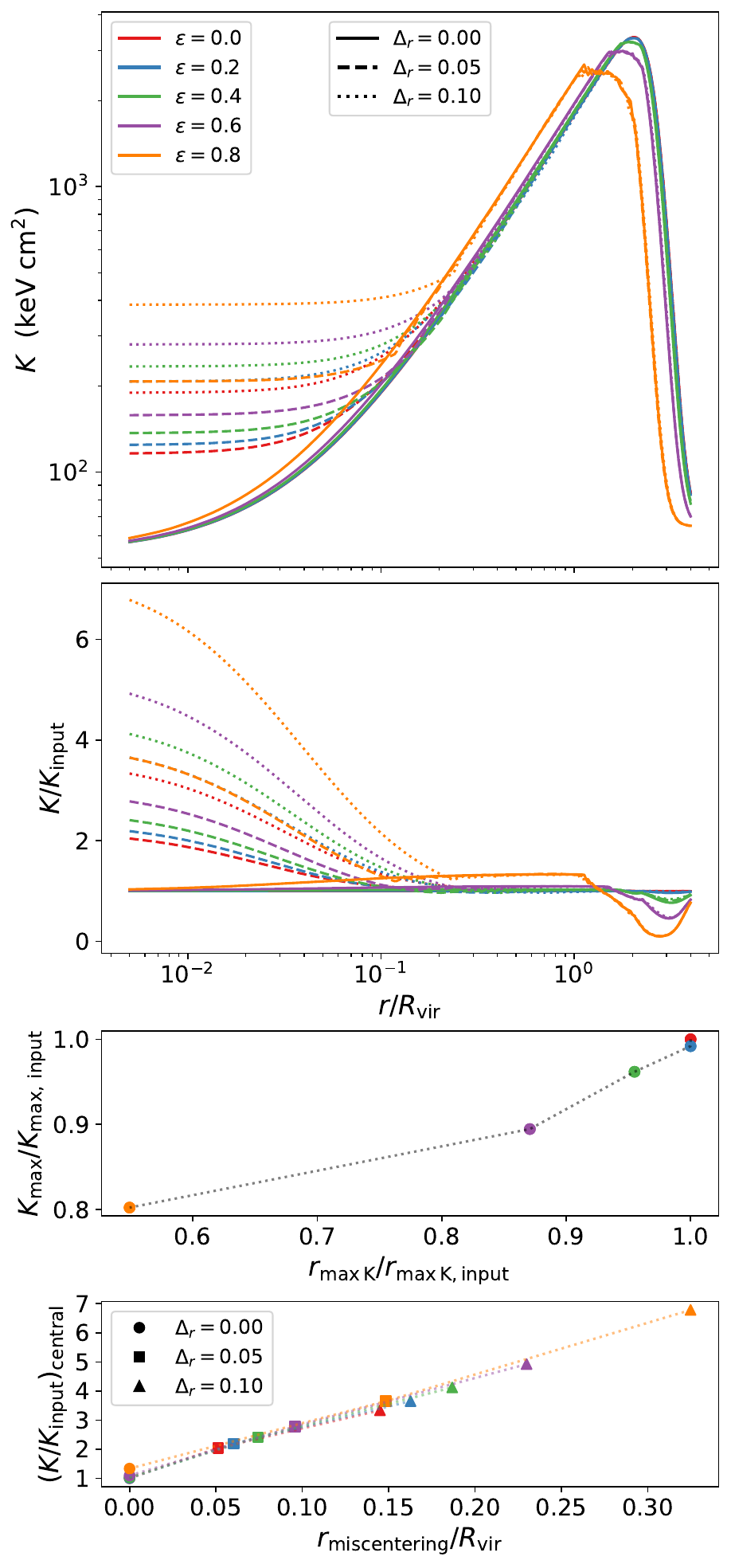}
    \caption{Density (left-hand side column) and entropy (right-hand side column) profiles in the mock test to assess the geometric effects of ellipticity and centre offset. \textit{Top panels:} profiles, colour-coded by the mock ellipticity value, the line style indicating $\Delta_r$. \textit{Second panels:} quotients between the input density and entropy profiles, and the ones recovered after the geometric transformations. \textit{Bottom panels:} subsequent panels summarise the effects of, both, ellipticity and centre offset, on flattening the density profiles, increasing the core entropy, and displacing the entropy peak in the outskirts. Here, $\Delta_r$ is represented by the marker shape.}
    \label{fig:mock_profiles}
\end{figure*}

This set-up allows us to study the purely geometric effect of ellipticity and an offset in the centre of the cluster on the density and entropy profiles, which we show, respectively, in the left-hand side and right-hand side panels of Fig. \ref{fig:mock_profiles}. The upper panels present the recovered profiles, with color hue indicating $\epsilon$ and lightness encoding $\Delta_r$. To better highlight the effects, the second panel in each column presents the quotient between these profiles and the input one (i.e. the solid, red line in the upper panels).

Regarding density profiles, with $\Delta_r = 0$ (solid lines), the effect of ellipticity on the median profiles is to reduce densities, especially in the outskirts. However, for realistic ICM ellipticity values ($\epsilon \ll 0.5$), this effect is contained ($\lesssim 10 \%$). Centre offset does play a crucial role in altering the recovered, spherical density profiles, as higher $\Delta_r$ considerably lowers central densities. In the lower-left panel we graphically summarise the minimum value of $\rho / \rho_\mathrm{input}$ and the \textit{miscentering radius}, $r_\mathrm{miscentering}$, which we define as the innermost point where the input and recovered density profiles start to diverge. This radius is typically in the range $r_\mathrm{miscentering} \in [1, 2] \Delta_r R_\mathrm{vir}$, and is higher for larger ellipticities.

For the entropy profiles, we observe a central entropy increase when $\Delta_r \neq 0$, that again is stronger for larger ellipticities, as a given radial shell will contain more material from outer elliptical shells. The magnitude of this central entropy boost can reach $\lesssim 4$ for realistic ICM ellipticities and offsets. Additionally, only for very high ellipticities, a tendency for the entropy peak to displace to slightly inner radii, and to decrease its magnitude (by $\sim 20\%$ at the unrealistically-high ellipticities). Overall, comparing with the effects seen in Fig. \ref{fig:entropy_indicators} in the main text, this geometrical effects are too weak and the trends we detect cannot be solely explained by geometry.

\begin{figure*}
    \centering
    \includegraphics[width=\linewidth]{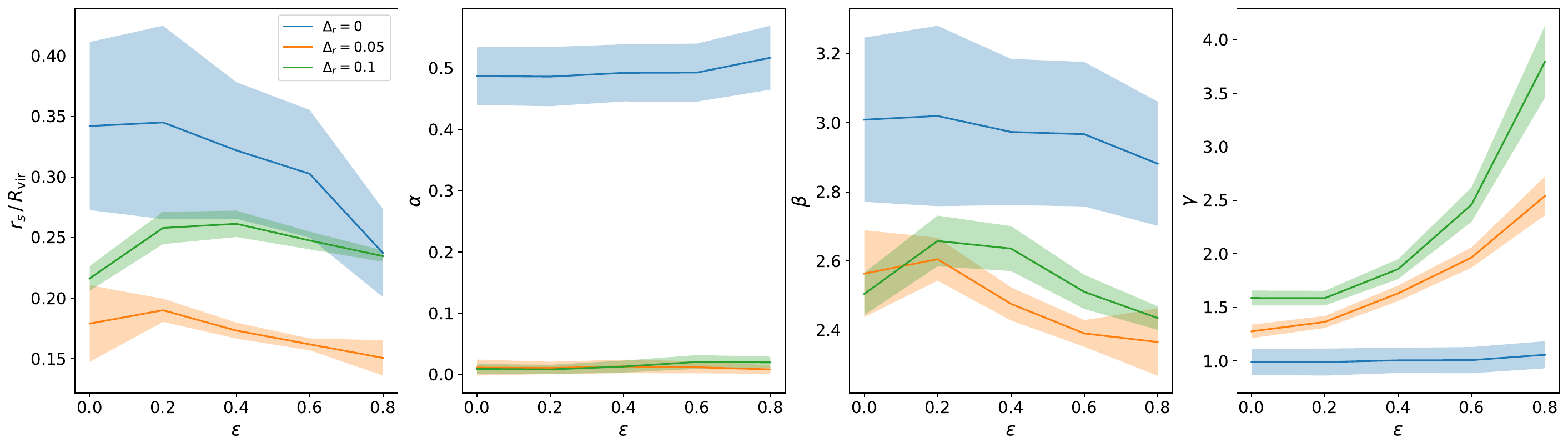}
    \caption{Effects of ellipticity (horizontal axis within each panel) and centre offset (line colours) on estimating the parameters of a gNFW profile. Left to right, the different panels show the scale radius in units of the virial radius, $r_s / R_\mathrm{vir}$, the inner slope $\alpha$, the outer slope $\beta$, and the transition sharpness parameter, $\gamma$.}
    \label{fig:mock_profiles_NFW}
\end{figure*}

These geometric effects on the ICM profiles computed assuming spherical symmetry have an effect on the estimation of certain parameters derived from the profiles. Although we have ruled out that the trends we observe in our simulation are solely due to geometry, we include this study here for completeness. Fig. \ref{fig:mock_profiles_NFW} shows how the scale radius (first panel) and the three slope parameters (subsequent panels) of the gNFW depend on ellipticity (horizontal axis) and centre offset (color) in the mock test. Somewhat surprisingly, the test shows a non-trivial (non-monotonic) relation between mock centre offset and $r_s$ (and, thus, NFW concentration). This is a consequence of the high correlation between $r_s$ and $\alpha$ (second panel). As soon as we slightly increase $\Delta_r$, the fit captures an inner slope $\alpha \approx 0$, which in turn alters the interpretation of the scale radius (as it no longer corresponds to the position at which the logarithmic slope is $-2$). Finally, the transition slope (fourth panel), $\gamma$, is strongly affected by both miscentering and ellipticity, as both these effects contribute to increasing anisotropy within the radial shells considered for computing the profiles and, therefore, smear out the transition.

\begin{figure*}
    \centering
    \includegraphics[width=0.8\linewidth]{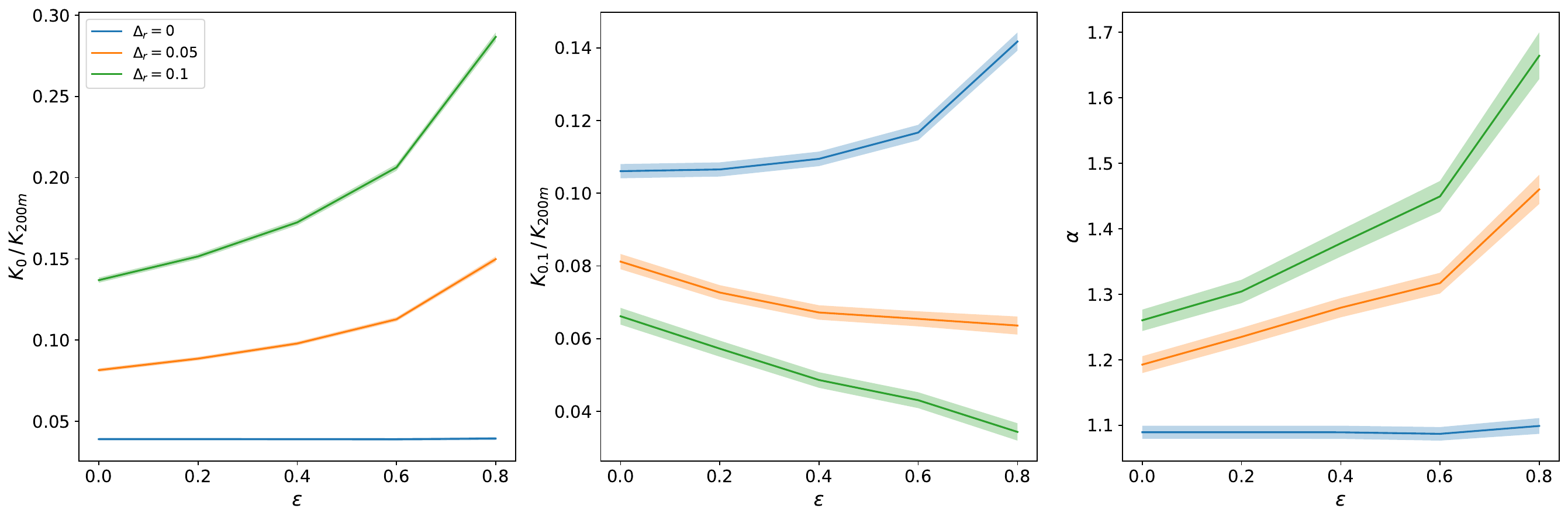}
    \caption{Effects of ellipticity (horizontal axis within each panel) and centre offset (line colours) on estimating the parameters of a \citet{Cavagnolo_2009} entropy profile. Left to right, the different panels show the core entropy excess, $K_0$; the power-law normalization at $0.1 R_{200m}$, $K_{0.1}$; and the power-law index, $\alpha$.} 
    \label{fig:mock_profiles_entropy}
\end{figure*}

We perform a similar analysis for the entropy profiles in Fig. \ref{fig:mock_profiles_entropy}. In this case, the trends are even cleaner, and suggest that ellipticity alone cannot alter the values of $K_0$ (left-hand side panel), but it can have a significant effect, boosting it by a factor of a few, in the presence of a non-null centre offset. Likewise, the slope of the power-law regime ($\alpha$) is significantly increased with ellipticity, for the non-null $\Delta_r$ case.

%--------------------------------------------------------------------
\section{Fitting details}
\label{app:fits}
%--------------------------------------------------------------------

The fits in Sect. \ref{s:results.fits} are carried out using an ensemble MCMC approach, implemented within the \texttt{emcee} package \citep{Foreman-Mackey_2013}. In particular, for each fit we initialise 100 chains, that are let to explore the posterior space for 2000 steps after an initial burn-in phase of 500 steps. The fits are performed assuming a chi-squared likelihood function in the logarithm of the target variable (e.g., $\log_{10} \rho$, $\log_{10} K$), where only the errors in the dependent variable are taken into account. Priors are flat within the bounds specified below for each case.

\paragraph{Fits for the NFW concentration.} The model function is 
\begin{equation}
    \rho(r) = \frac{\rho_0}{\frac{r}{r_s} \left( 1 + \frac{r}{r_s}  \right)^2},
    \label{eq:NFWform}
\end{equation}
\noindent where we adopt flat priors on $\log_{10} (\rho_0/\rho_B) \in [0, 8]$ and ${r_s/R_\mathrm{vir} \in [0.01, 1]}$. The fits are restricted to the radial range where $\rho / \rho_B > 1$.

\paragraph{Fits to the gNFW functional form.} For the generalised NFW profile, we consider the functional form
\begin{equation}
    \rho(r) = \frac{\rho_0 \, \cdot \, 2^\frac{\beta-\alpha}{\gamma}}{\left(\frac{r}{r_s}\right)^{\alpha} \left[ 1 + \left(\frac{r}{r_s}\right)^\gamma  \right]^\frac{\beta-\alpha}{\gamma}},
    \label{eq:gNFWform}
\end{equation}
\noindent where the power of two term is added to remove the covariance of $\rho_0$ with the slope parameters (so that $\rho_0$ explicitly represents the normalisation of the profile at $r_s$). While the density and radius normalisation parameters keep the same priors, for the slopes we consider the bounds $\alpha, \, \beta \in [0, 10]$, $\gamma \in [0.2, 5]$. The fits are restricted to the radial range where $\rho / \rho_B > 1$ ($P / P_{200m} > 1$ for the pressure profiles).

\paragraph{Fits to the entropy profile.} For the entropy profile, we adopt the functional form by \citet{Donahue_2006} and \citet{Cavagnolo_2009}, modified to normalise the power-law at $0.1 R_\mathrm{200m}$ instead of a constant $100 \, \mathrm{kpc}$,
\begin{equation}
    K(r) = K_0 + K_{0.1} \left( \frac{r}{0.1 R_{200m}} \right)^\alpha,
    \label{eq:cavagnolo}
\end{equation}
\noindent where, in this case, the priors are flat in $K_0 / K_{200m} \in [0, 1]$, $K_{0.1} / K_{200m} \in [0, 10]$, and $\alpha \in [0, 5]$. To make the results representative of the typical self-similar range of entropy profiles (see lower panel of Fig. \ref{fig:scatter}), and mitigate the effects of overcooling in the centre and the accretion shocks in the outskirts, the likelihood function is computed only over the radial range $0.1 \leq r / R_{200m} \leq 2$.

\paragraph{Polynomial fits.} The fits for summarising the trends of the profile fit parameters with accretion rate are not performed by MCMC sampling, but instead we just proceed with an ordinary least squares fit. The process for deciding the degree of the fitting polynomial is analogous to the one described in \citet{Valles-Perez_2023}. However, here we also account for the errors in the independent variable, by estimating the local slope of the fitting function at each point and iterating the process. This is crucial in this case, since the width of the bins in $\Gamma_{200m}$ is highly variable.

\end{document}